\documentclass[%
 reprint,
 groupedaddress,
 amsmath,amssymb,
 aps,
 pra
]{revtex4-1}

\usepackage[pdftex]{graphicx, color}
\usepackage[pdftex]{hyperref}
\usepackage{bm}
\usepackage{braket}
\begin{document}
\title{Impact of the Rashba Spin Orbit Coupling on $f$-electron Materials}

\author{Yoshihiro Michishita}
\affiliation{Department of Physics, Kyoto University, Kyoto 606-8502, Japan}
\author{Robert Peters}
\affiliation{Department of Physics, Kyoto University, Kyoto 606-8502, Japan}
\begin{abstract}
The combination of strong spin orbit coupling and strong correlations holds tremendous potential for interesting physical phenomena as well as applications in spintronics and quantum computation.
In this context, we here study the interplay between the Rashba spin-orbit coupling (RSOC) and the Kondo screening in noncentrosymmetric $f$-electron materials.
We show that the Kondo coupling of the $f$-electrons becomes anisotropic at high temperatures due to the RSOC. 
However, an isotropic Kondo effect is restored at low temperature which leads to a complete Kondo screening.
We furthermore demonstrate that the Kondo effect has influence on the Rashba splitting in the band structure, which becomes temperature dependent. 
Although the $f$-electrons are localized at high temperature, a helical spin polarization of the conduction band emerges due to the scattering with the $f$-electrons.
With decreasing temperature, the Kondo screening occurs, which leads to drastic changes in the band structure. Remarkably, these changes in the band structure depend on the helical spin polarization. For strong RSOC, we observe that the hybridization gap of one of the helical bands is closed at low temperature and a helical half-metal is formed.
\end{abstract}

\newcommand{\cm}[1]{}
\newcommand{\cmm}[1]{{\color{red}\sout{ #1}}}

\maketitle
\section{Introduction}
Recently, the combination of strong spin orbit coupling (SOC) and strong correlations have aroused great interest\cite{annurev-conmatphys-020911-125138}.
For light elements, SOC has usually only a weak effect on the material properties. However, in heavy elements, such as in $f$-electron materials, SOC can become large and strongly affects the low temperature properties\cite{PhysRevB.97.115128}.  Moreover, in materials with partially filled $f$-electron orbitals, strong electron correlations can commonly be observed, which results in intriguing physics such as magnetism, quantum criticality, and unconventional superconductivity\cite{coleman2006heavy}.
The combination of both, strong correlations and strong spin orbit interaction, can thus lead to completely novel phenomena such as spin-orbit assisted Mott insulators \cite{PhysRevLett.101.076402}, spin liquids \cite{annurev-conmatphys-020911-125138} and correlated topological insulators such as SmB$_6$ or YbB$_{12}$\cite{neupane2013surface,PhysRevLett.112.016403,PhysRevLett.110.096401,doi:10.1146/annurev-conmatphys-031214-014749,PhysRevLett.104.106408}.

Another intriguing aspect arises when the inversion symmetry in these materials is broken. This results in the appearance of the anti-symmetric spin orbit coupling (ASOC)\cite{manchon2015new}, such as the Rashba SOC (RSOC). ASOC leads to a spin-momentum locking of the electrons, which makes it possible to manipulate the magnetization or the spin direction of the electrons by applying electric fields, which might be used in spintronics devices such as a spin transistor or spin-orbit qubits.\cite{manchon2015new}
ASOC also plays an important role in the magneto electric effect\cite{PhysRevB.87.174411}, the switching of the spin texture\cite{PhysRevLett.102.017205} and  topologically nontrivial band structure which results in the quantum spin hall effect\cite{PhysRevLett.104.126802}.
Therefore, the combination of ASOC and strong correlations can be expected to yield fascinating phenomena\cite{fujimoto2012aspects}.
Noncentrosymmetric $f$-electron materials, which include strong correlations and might include strong RSOC, are thus good candidates for studying this combination. Examples of these materials are CePt$_3$Si, CeRhSi$_3$ and CeIrSi$_3$\cite{PhysRevLett.92.027003,PhysRevLett.95.247004,sugitani2006pressure}.

Besides these noncentrosymmetric $f$-electron materials, recently, $f$-electron superlattices have been proposed as a new platform for utilizing the combination of strong RSOC and strong correlations\cite{PhysRevLett.112.156404,petrovic2001heavy,PhysRevLett.94.057002,goh2012anomalous,PhysRevB.94.205142}. 
In these $f$-electron superlattices, the inversion symmetry is broken around the interface between different materials. Experiments have demonstrated that the strength of the RSOC depends on the structure of the superlattice, i.e. how many atomic layers of one material are used\cite{PhysRevLett.112.156404}. Thus, the strength of the RSOC seems to be controllable, which might give a platform to study the interplay between strong correlations and strong RSOC in a controlled fashion. 

To understand these phenomena seen in these superlattices and noncentrosymmetric $f$-electron materials, it is necessary to understand the nature of the quantum state arising from the interplay between strong Coulomb interaction and RSOC.
The Coulomb interaction in $f$-electron materials leads to the Kondo effect\cite{coleman2006heavy,kondo1964resistance}. At high temperature, the $f$-electrons form localized magnetic moments, which are screened by the conduction electrons at the Kondo temperature. Due to this screening, $f$-electrons become itinerant and participate in the Fermi surface, forming a so-called heavy-fermion state.
On the other hand, the RSOC induces an anisotropic spin splitting into the band structure, which can be expected to hinder the formation of magnetic moments,
 thus competing with the Kondo effect. Both energy scales do not need to be small, as the example of CePt$_3$Si shows, for which a Kondo temperature of $T_K = 80K$ \cite{PhysRevLett.92.027003} and the spin-orbit interaction of $50meV-200meV$ \cite{PhysRevB.69.094514} have been determined.
 
 Until now, theoretical works have focused on the impact of the RSOC on the Kondo effect in quantum dots and impurity models. Depending on the impurity model, there are predictions that the Kondo effect is enhanced exponentially\cite{PhysRevLett.108.046601}, is almost unchanged\cite{0953-8984-28-39-396005}, or is suppressed\cite{vernek2009kondo,misiorny2011interplay}  by the RSOC.

 However, the effect of RSOC on the properties of $f$-electron materials, where every atom possesses a magnetic moment due to a partially filled $f$-orbital, is poorly studied. The Kondo effect occurs in these materials not as a screening of an isolated magnetic moment embedded into a conduction band, but the Kondo effect occurs coherently in all magnetic moments. 

 In this work, we study the interplay between the RSOC and the Kondo effect in the half-filled periodic Anderson model (PAM).
In the first part, we analyze the impact of the RSOC on the Kondo effect by using perturbation theory and the
dynamical mean field theory (DMFT) combined with the numerical renormalization group (NRG)\cite{RevModPhys.68.13,RevModPhys.80.395,PhysRevB.74.245114}. By looking from two sides on the same problem, we can understand how $f$-electrons are screened by the Kondo effect in noncentrosymmetric materials.
In the second part, we use DMFT to analyze the impact of the Kondo effect on the Rashba splitting in the band structure.
Remarkably, we find that the hybridization strength between $c$- and $f$-electrons depends on the helical polarization of the band, which has a strong impact on the temperature dependence of the band structure.

The rest of the paper is organized as follows:
 In Sec. \ref{intro}, we introduce the model Hamiltonian that we use in this paper. In Sec. \ref{ana_Tk}, we derive an effective Hamiltonian (Kondo lattice model) by second order perturbation theory and calculate the Kondo temperature by poor man's scaling. 
Then, by using DMFT combined with NRG, we calculate the temperature-dependent magnetic moment of the $f$-electrons from which we can numerically deduce the Kondo screening and Kondo temperature.
Finally, in Sec. \ref{Kondo_on_Rashba} we analyze the temperature dependent band structure by studying the spectral functions and the conductivity for both helical spin polarizations.

 \section {Model and Methods}\label{intro}
 Because atoms with partially filled $f$-electron orbitals are generally large, the SOC can be expected to be strong in these materials. To analyze the interplay between the RSOC and the Kondo effect, we use a periodic Anderson model where the SOC is generated only by the $f$-electron band, resulting in a RSOC within the $f$-electron band and between the $c$- and the $f$-electrons.
 An analysis of the RSOC within the $c$-electron band is left as a future study. 
 Our Hamiltonian, which is based on a model of CePt$_3$Si\cite{yanase2008superconductivity}, reads
\begin{align}
\thickmuskip=0mu
\medmuskip=0mu
\thinmuskip=0mu
&{\cal H}\mathalpha{=}\!\sum_{\bm{k,\sigma,\sigma^{\prime}}}\Bigl(\epsilon_{\bm{k}}\sigma^0c^{\dagger}_{\bm{k}\sigma}c_{\bm{k}\sigma^{\prime}}\nonumber\\
 &\mathalpha{+} (\epsilon_{f\bm{k}}\sigma^0 \mathalpha{+} \alpha_{ff}(\sigma^x\sin{k_y} \mathalpha{-} \sigma^y\sin{k_x}))_{\sigma\sigma^{\prime}}f^{\dagger}_{\bm{k}\sigma}f_{\bm{k}\sigma^{\prime}}\nonumber\\
&\mathalpha{+}(V\sigma^0 \mathalpha{+} \alpha_{cf}(\sigma^x\!\sin{k_y} \mathalpha{-} \sigma^y\!\sin{k_x}))_{\sigma\sigma^{\prime}}(f^{\dagger}_{\bm{k}\sigma}c_{\bm{k}\sigma^{\prime}}\mathalpha{+}h.c)\Bigr)\nonumber\\
&\mathalpha{+}U \sum_i n_{i\uparrow}n_{i\downarrow}\label{X}\\
&\epsilon_{\bm{k}}=2t_c (\cos{k_x}+\cos{k_y})+\mu_c\\
&\epsilon_{f\bm{k}}=2t_f (\cos{k_x}+\cos{k_y})+\mu_f
\end{align}
where $c^{(\dagger)}_{\bm{k}\sigma},f^{(\dagger)}_{\bm{k}\sigma}$ are annihilation (creation) operators of the conduction and the $f$-electrons for momentum $\bm{k}$ and spin-direction $\sigma$. $t_{c,f}$ are the inter-site hopping strengths for the $c$- and $f$-electrons. For simplicity  we assume a two-dimensional square lattice.
$\mu_{c,f}$ are the chemical potentials for the $c$- and the $f$-orbitals. $V$ describes a local hybridization between the $c$- and the $f$-orbitals, $\alpha_{ff}$ is the RSOC within the $f$ electron band, and $\alpha_{cf}$ is the RSOC between the $c$- and the $f$-electron bands. 
Throughout this paper, we fix $t_f=- 0.05 t_c$ and use $t_c=1$ as unit of the energy.

Besides analytical tools like perturbation theory, we employ DMFT combined with NRG to study this Hamiltonian. DMFT takes local fluctuations fully into account by self-consistently solving the mean field equations\cite{RevModPhys.68.13}. The lattice Hamiltonian is thereby mapped onto a quantum impurity model. Thus, DMFT neglects nonlocal fluctuations which becomes exact in infinite dimensions. To solve the quantum impurity model, we use the NRG, which calculates low energy properties by iteratively  discarding high-energy states. It has been shown that NRG is a very reliable tool at low temperature\cite{RevModPhys.80.395,PhysRevB.74.245114}.

\section {Effect of the Rashba coupling on the Kondo effect}
\label{ana_Tk}

First, we study the effect of the RSOC on the Kondo screening. For this purpose, we use two different techniques, namely perturbation theory and DMFT.
By using different techniques, we can analyze different aspects of the impact of the RSOC on the Kondo screening. While perturbation theory is only correct in the limit of very strong interactions (weak hybridization), it includes the momentum dependence of the RSOC. On the other hand, DMFT is a reasonable approximation for any interaction strength and can describe the crossover from localized to itinerant $f$-electrons, but it neglects the momentum dependence of the self-energy.

\subsection {Kondo screening analyzed by perturbation theory}
By using perturbation theory and poor man's scaling, we can analyze the impact of the RSOC on the Kondo screening as long as the $f$-electrons can be regarded as localized.
We first derive an effective Kondo lattice model (KLM), by treating the $c$-$f$ hybridization in  second order perturbation theory for
$U \gg t,V,\alpha$ where the ground state is half-filled. We therefore fix the chemical potential of the $f$-electrons as $\mu_f= - U/2$, and derive the Kondo coupling for $\alpha_{cf} \neq 0$ and $\alpha_{ff}=0$. We focus here on this situation, because $\alpha_{cf}$ dominantly influences the Kondo effect. The effect of $\alpha_{ff}$ and further details of the derivation are given in appendix \ref{APP_KLM}.

The effective Hamiltonian becomes
\begin{align}
&{\cal H}_{KLM}\mathalpha{=}\!\sum_{\bm{k}}\epsilon_{\bm{k}}c^{\dagger}_{\bm{k}\sigma}c_{\bm{k}\sigma}\\ \mathalpha{+} 
&\sum_{i}\sum_{p=\{x,y,z\}} \sum_{\bm{k} \bm{k^{\prime}}} \exp^{i (\bm{k-k^{\prime}}) \cdot \bm{r}_i} J^{p}_{\bm{k k^{\prime}}} S^p_i c^{\dagger}_{\bm{k^{\prime}}} \sigma^{p} c_{\bm{k}} \\
&J^x_{\bm{k k^{\prime}}}\mathalpha{=} 2(V^2-\alpha^2_{cf}(\sin{k_x} \sin{k^{\prime}_x}-\sin{k_y} \sin{k^{\prime}_y}))/U\\
&J^y_{\bm{k k^{\prime}}}\mathalpha{=} 2(V^2-\alpha^2_{cf}(-\sin{k_x} \sin{k^{\prime}_x}+\sin{k_y} \sin{k^{\prime}_y}))/U\\
&J^z_{\bm{k k^{\prime}}}\mathalpha{=} 2(V^2-\alpha^2_{cf}(\sin{k_y} \sin{k^{\prime}_y}+\sin{k_x} \sin{k^{\prime}_x}))/U.
\end{align}
Contrary to the ordinary Kondo lattice model, the coupling between the localized spins and the $c$ electrons is strongly momentum dependent. Moreover, the coupling is anisotropic, i.e. $J^x\neq J^y\neq J^z$. Such a Kondo coupling can lead to a Kondo singlet with internal angular momentum, e.g. $d$-wave symmetry\cite{PhysRevB.75.144412}, which means that the Kondo singlet becomes nonlocal\cite{PhysRevB.77.125118}.

As a first start to analyze the impact of the RSOC on the Kondo effect, we here constrain our results to the conventional Kondo singlet and nonmagnetic heavy Fermion phase with $s$-wave symmetry, where we can assume that the scattering close to the Fermi surface of the unperturbed system with $\bm{q}=\bm{k}^{\prime}-\bm{k}\simeq0$ is dominant.
We obtain the simplified Kondo lattice model,

\begin{align}
&{\cal H}_{KLM}\nonumber\\
& \ \mathalpha{\simeq}\!\sum_{\bm{k}}\epsilon_{\bm{k}}c^{\dagger}_{\bm{k}\sigma}c_{\bm{k}\sigma} \mathalpha{+} \sum_{i,\bm{k},\bm{q}\simeq0} (J_0 \bm{S_i} \cdot (c^{\dagger}_{\bm{k+q}} \bm{\sigma} c_{\bm{k}}) -J_{R} S^z_i c^{\dagger}_{\bm{k+q}} \sigma^{z} c_{\bm{k}})\label{1}\\
& J_0 = \frac{2V^2}{U}\label{2}\\
& J_{R} = \frac{1}{4 \pi^2} \sum_{\bm{k}}\frac{2\alpha^2_{cf}(\sin^2{k_x}+\sin^2{k_y} )}{U} = \frac{2\alpha^2_{cf}}{U} \label{3},
\end{align}
where $J_R$ is the spin exchange interaction generated by the RSOC. The main difference to the ordinary Kondo lattice model is an anisotropic exchange interaction arising from $J_R$.

We can now use poor man's scaling to integrate out high energy states and derive the Kondo temperature\cite{PhysRevLett.40.416}. The Renormalization group (RG) equations for the coupling parameters $J^z, J^{\parallel}$ are
\begin{eqnarray}
\frac{dJ_z}{dE} = - \frac{n(\epsilon_F) J^2_{\parallel}}{E} \label{renorm_z}\\
\frac{dJ_{\parallel}}{dE} = - \frac{n(\epsilon_F) J_z J_{\parallel}}{E}\label{renorm_p}
\end{eqnarray}

which satisfy
\begin{eqnarray}
J^2_{\parallel} - J^{2}_z = J_{R}( - J_{R} +2J) = a^2 (const.)\label{bind}
\end{eqnarray}

\begin{figure}
\includegraphics[
height=2.0in,
width=3.0in
]{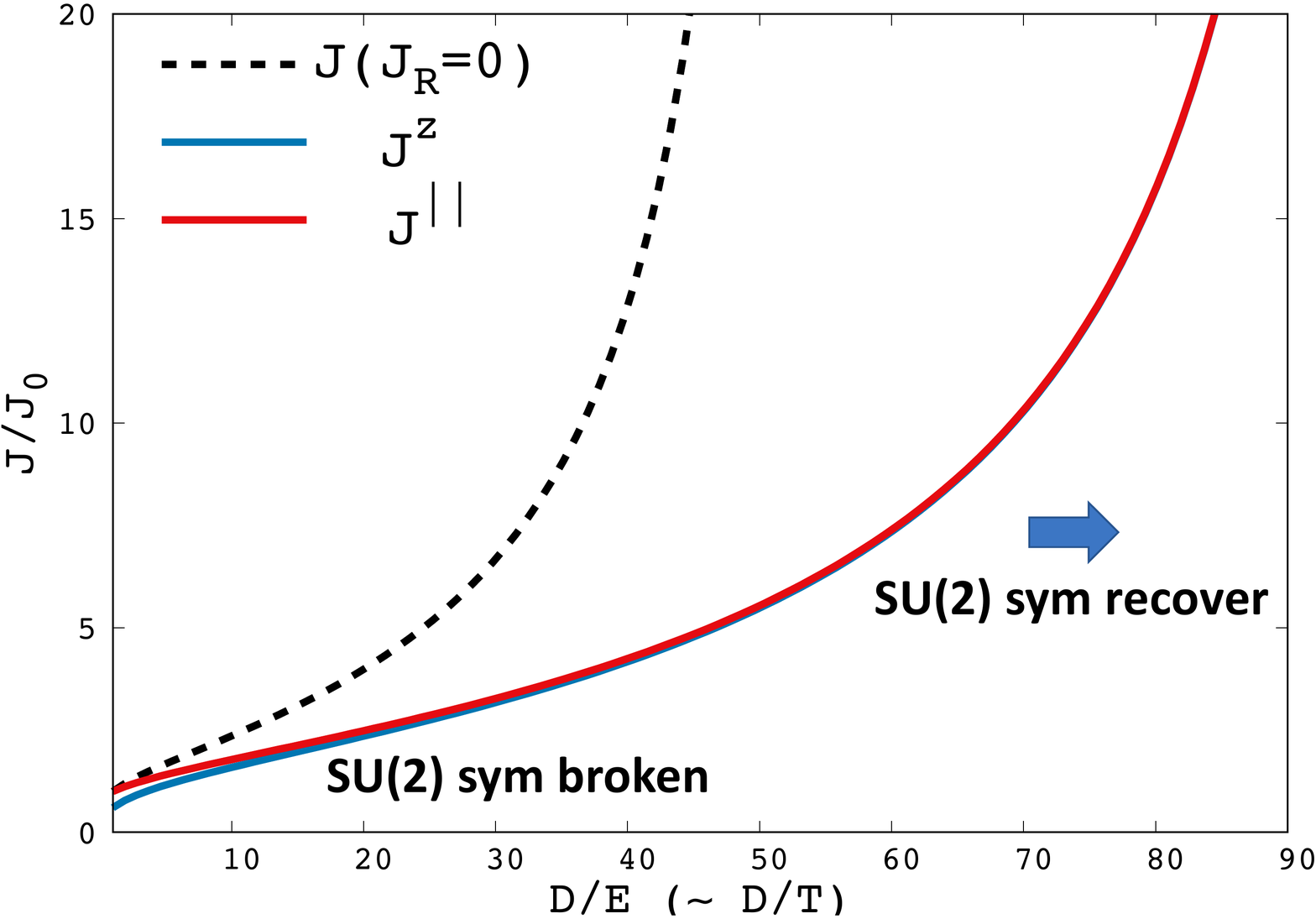}
\caption{Renormalization flow of $J^z$ and $J^{||}$.
The dashed line corresponds to the centrosymmetric case, with $J_0=2V^2/U=0.05$. The parameters in the noncentrosymmetric case are $J_0=2V^2/U=0.05$, $J_R=2\alpha^2_{cf}/U=0.02$.}
\label{fig:T_K2}
\end{figure}
\begin{figure}
\includegraphics[
height=2.0in,
width=3.0in
]{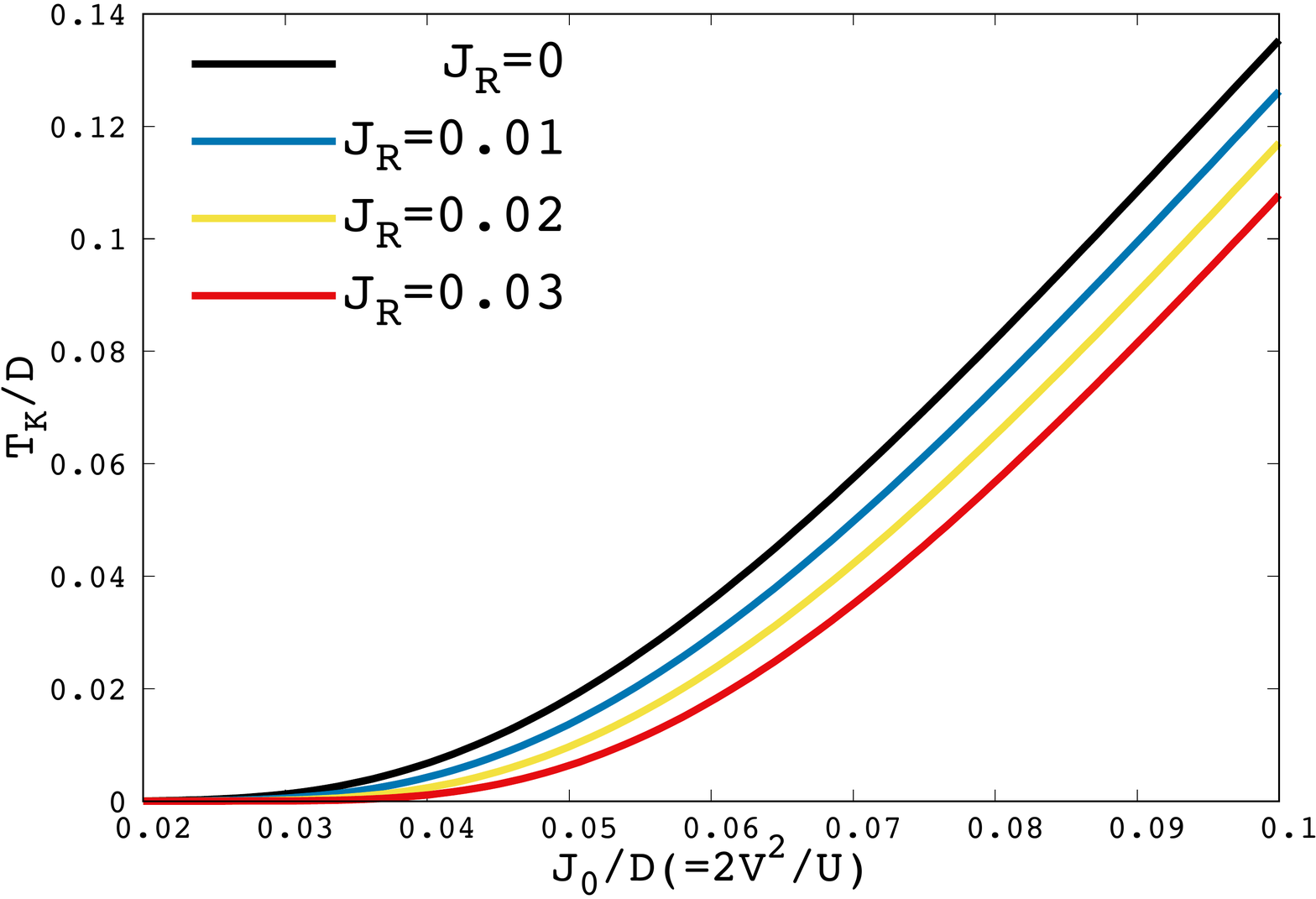}
\caption{Kondo temperature $T_K$  calculated by scaling theory in the anisotropic Kondo model for $\alpha_{cf} \neq 0,\alpha_{ff}=0$\label{fig:T_K}}
\end{figure}

From Eq. (\ref{renorm_z}), (\ref{renorm_p}), and (\ref{bind}), we derive two independent RG equations,
\begin{eqnarray}
\frac{dJ_z}{dE} = - \frac{n(\epsilon_F) (J^2_{z}+a^2)}{E}  \\
\frac{dJ_{\parallel}}{dE} = - \frac{n(\epsilon_F) J_{\parallel} \sqrt{J^2_{\parallel} - a^2} }{E}.
\end{eqnarray}
From  these two RG equations we can find the energy scales $T^z, T^{\parallel}$ for  $J^{z(\parallel)}$, which are given as 
\begin{eqnarray}
T^z_K &=& t_c \exp [- \arctan{(a/(J_0-J_R))}/(n(\epsilon_F) a)] \label{Tk_z}\\
T^{\parallel}_K &=& t_c \exp [- (\arcsin{(a/J_0)})/(n(\epsilon_F) a)]\\
&=&T^z_K \label{Tk_p}\\
&\simeq& \exp [- 1/(n(\epsilon_F) J_0)] \times \exp [-a^2/(6J^{3})]
\end{eqnarray}
These results are consistent with the ordinary Kondo temperature for $\alpha_{cf}= \alpha_{ff} = 0$.

  By using perturbation theory, we see that the RSOC  leads to anisotropically coupled local moments, see Eq. (\ref{1}), as shown by the Schrieffer-Wolff transformation\cite{PhysRevLett.108.046601}.
    However, the screening occurs in such a way that at the Kondo temperature an isotropic singlet is formed; in the scaling theory, both coupling parameters diverge at the same Kondo temperature.
We show in Fig. \ref{fig:T_K2} the flow of the coupling parameters $J^z$ and $J^\parallel$, which diverge at the Kondo temperature. We see that although the coupling parameters are different at the beginning of the scaling, $D/E=1$, this difference vanishes during the scaling. Thus, a SU(2) symmetric Kondo singlet is formed at the Kondo temperature. 
The Kondo temperatures for different $J_R$ as calculated by scaling theory are shown  in Fig. \ref{fig:T_K}. We see that the anisotropy suppresses the Kondo temperature depending on $J_R$($\propto \alpha^2_{cf}/U$).

\subsection {Kondo screening analyzed by DMFT}

\begin{figure}
 \includegraphics[
height=2.2in,
width=3.2in
]{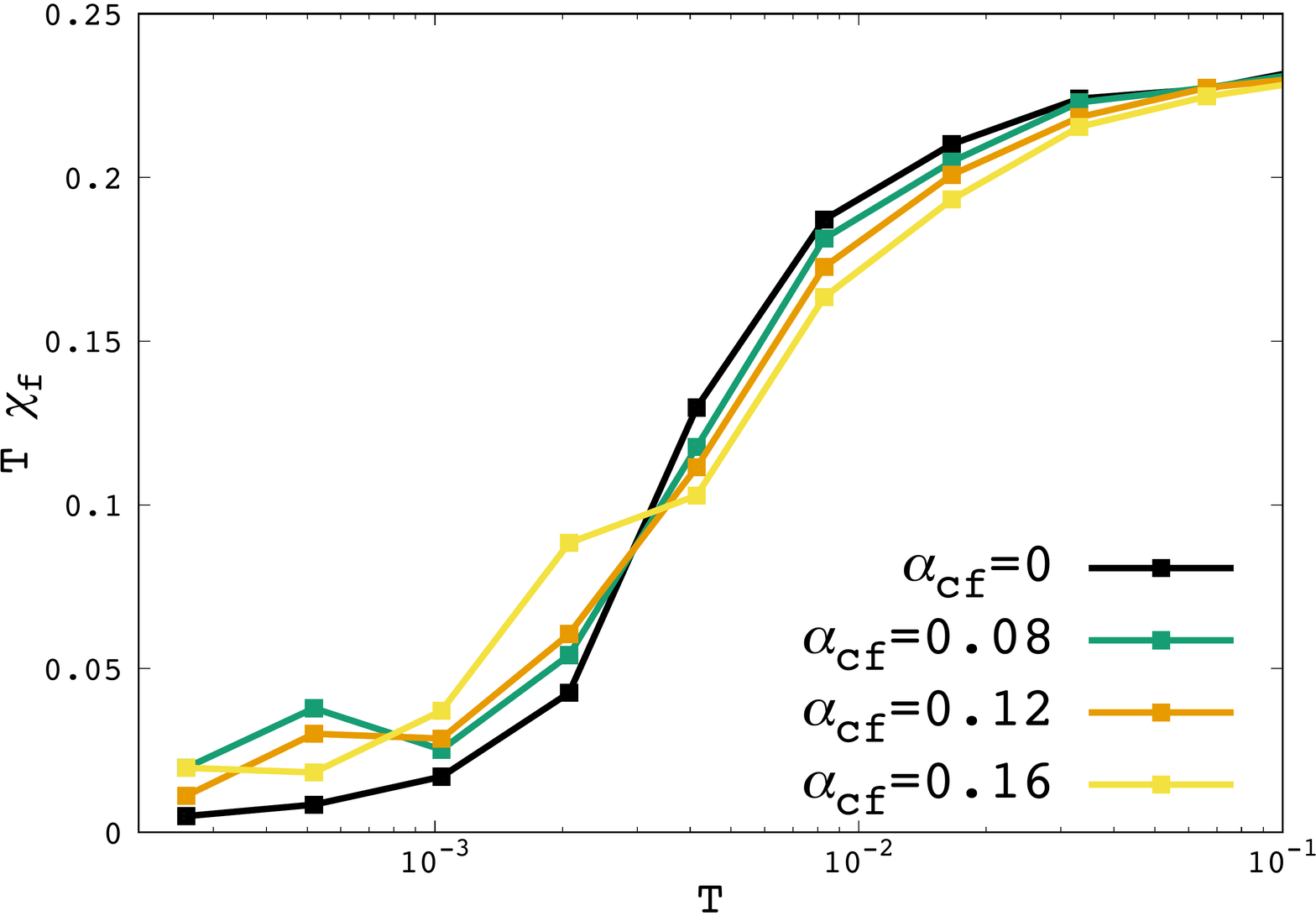}
\caption{Local magnetic moment, $T\chi_f$, calculated by DMFT for $\alpha_{cf} \neq 0,\alpha_{ff}=0$ and
 $t_c$=1.0, $t_f$=-0.05, $V$=0.4, $U$=2.0, $\mu_c$=0, $\mu_f$=-1.0.}\label{fig:Moment2}
 \includegraphics[
height=2.2in,
width=3.2in
]{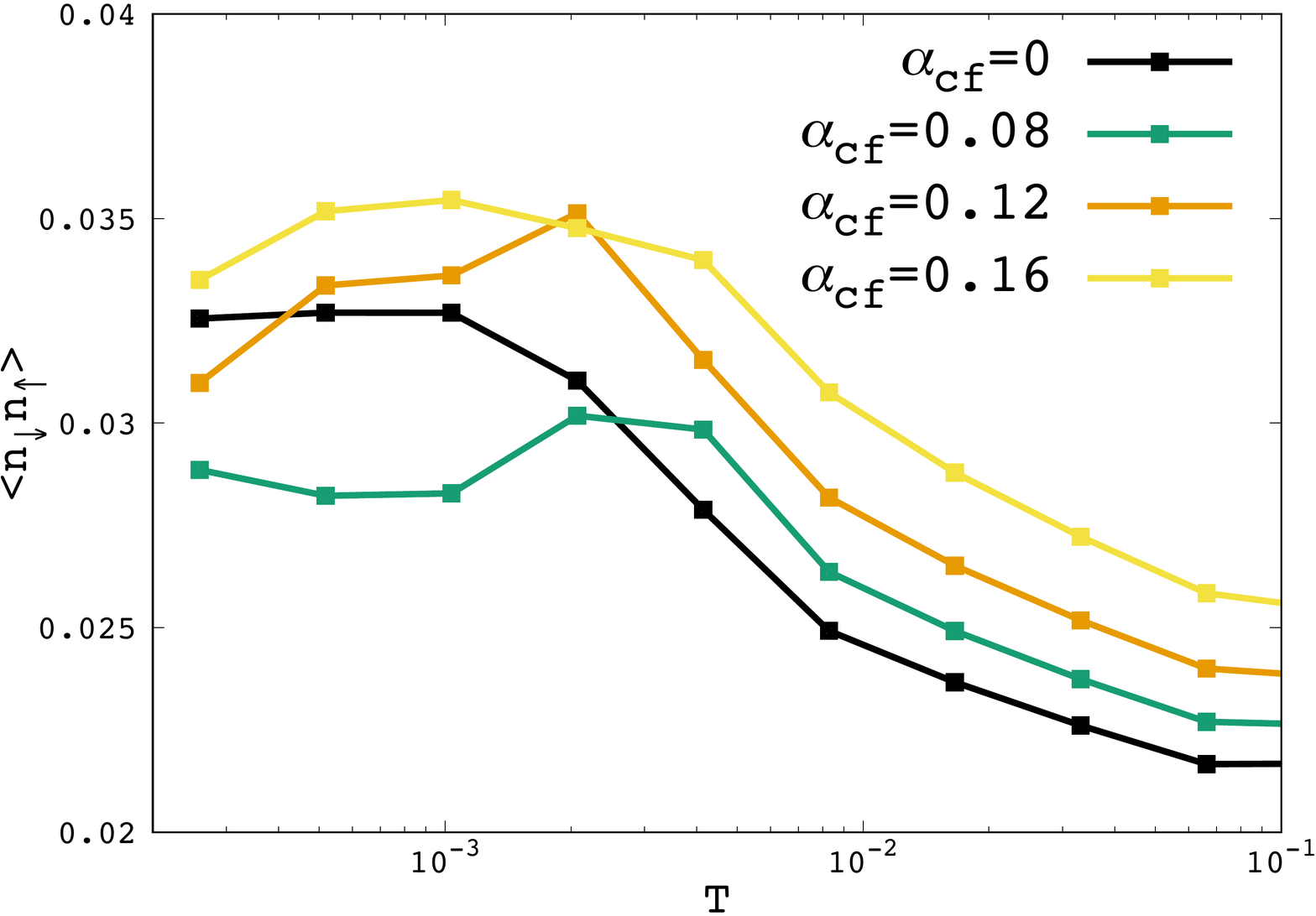}
\caption{Double occupancy of the $f$ electrons calculated by DMFT.
 The parameters are the same as in Fig. \ref{fig:Moment2}.}
\label{fig:DO}
\end{figure}

Up to now, we have studied the impact of the RSOC on the Kondo effect by scaling theory for an effective anisotropic Kondo  model.
The calculations have shown a suppression of the Kondo temperature due to the RSOC. However, this method gives only limited information about the strongly correlated quantum state in the periodic $f$-electron model in Eq.(\ref{X}). 
We now analyze the impact of the RSOC on the Kondo screening in noncentrosymmetric $f$-electron systems by using DMFT.  
Although DMFT takes the lattice structure into account and also includes local correlations exactly, we note that DMFT neglects the SU(2)-symmetry breaking generated by the RSOC. 
When calculating the local Green's function, which is the main ingredient for the self-consistent DMFT calculation, the momentum integration results in an artificial restoration of the SU(2) symmetry of the interaction. Instead of an anisotropic coupling between the $c$- and the $f$-electrons, the coupling becomes isotropic due to the DMFT approximation.
Thus,  DMFT calculations need to be scrutinized particularly at high temperatures, where the coupling between the $c$- and the $f$-electrons is anisotropic. 
 A more accurate analysis at high temperatures using cluster DMFT, which can take the SU(2) symmetry breaking into account, is left as a future study.

In Fig. \ref{fig:Moment2}, we show the local contribution of the $f$-electrons to the magnetic susceptibility $T\chi^z_f(T)$ calculated by using DMFT/NRG.
The magnetic susceptibility is determined by calculating the magnetic polarization of the $f$-electrons in a system with an applied small magnetic field.
This quantity shows a drastic change due to the occurring Kondo screening. $T\chi^z_f$ changes from $0.25$ at high temperatures, indicating a free spin $1/2$ to $0$ at low temperatures indicating completely screened $f$-electrons.

We observe that the screening of the magnetic moments starts at a higher temperatures when the RSOC becomes finite, $\alpha_{cf}\neq 0$. 
The RSOC $\alpha_{cf}$ acts as an additional hybridization between the $f$- and the $c$-electrons.
However, although the Kondo screening starts at higher temperatures, we observe that the magnetic moment, $T\chi_f$, is enhanced at low temperatures compared to $\alpha_{cf}=0$. The temperature at which $T\chi^z_f(T)$ vanishes is actually suppressed by the RSOC. Thus, the formation of the Kondo singlet occurs at a lower temperature in the noncentrosymmetric system, which agrees with scaling theory.

The crossover from localized to itinerant $f$-electrons can also be observed in the double occupancy of the $f$-electrons, $\langle n_\uparrow n_\downarrow\rangle$, see Fig. \ref{fig:DO}. 
When the Kondo screening starts at high temperatures, the $f$-electrons hybridize with the conduction electrons and become itinerant so that the double occupancy of the $f$-electrons increases. 
In the centrosymmetric system, $\alpha_{cf}=0$, the double occupancy increases with decreasing temperature and saturates, which corresponds to the completion of the Kondo screening. For $\alpha_{cf}\neq 0$, the double occupancy reaches a peak at intermediate temperatures and starts to decrease again with decreasing temperature. Thus, the double occupancy saturates at a lower temperature compared to the centrosymmetric system.
As in perturbation theory, the ground state is reached at a lower temperature in the noncentrosymmetric system. 

We thus reach to the following picture of the Kondo screening in noncentrosymmetric systems: The screening starts at higher temperature compared to the centrosymmetric material. At this temperature, the coupling between the $c$- and the $f$-electrons is anisotropic. However, when lowering the temperature, an SU(2) symmetric Kondo screening is restored and a complete Kondo singlet is formed at low temperatures. The temperature, at which the Kondo screening is completed, is suppressed by the RSOC.

\section{ impact of the Kondo effect on the Rashba splitting}
\label{Kondo_on_Rashba}
\begin{figure*}[t]
\includegraphics[width=0.3\linewidth]{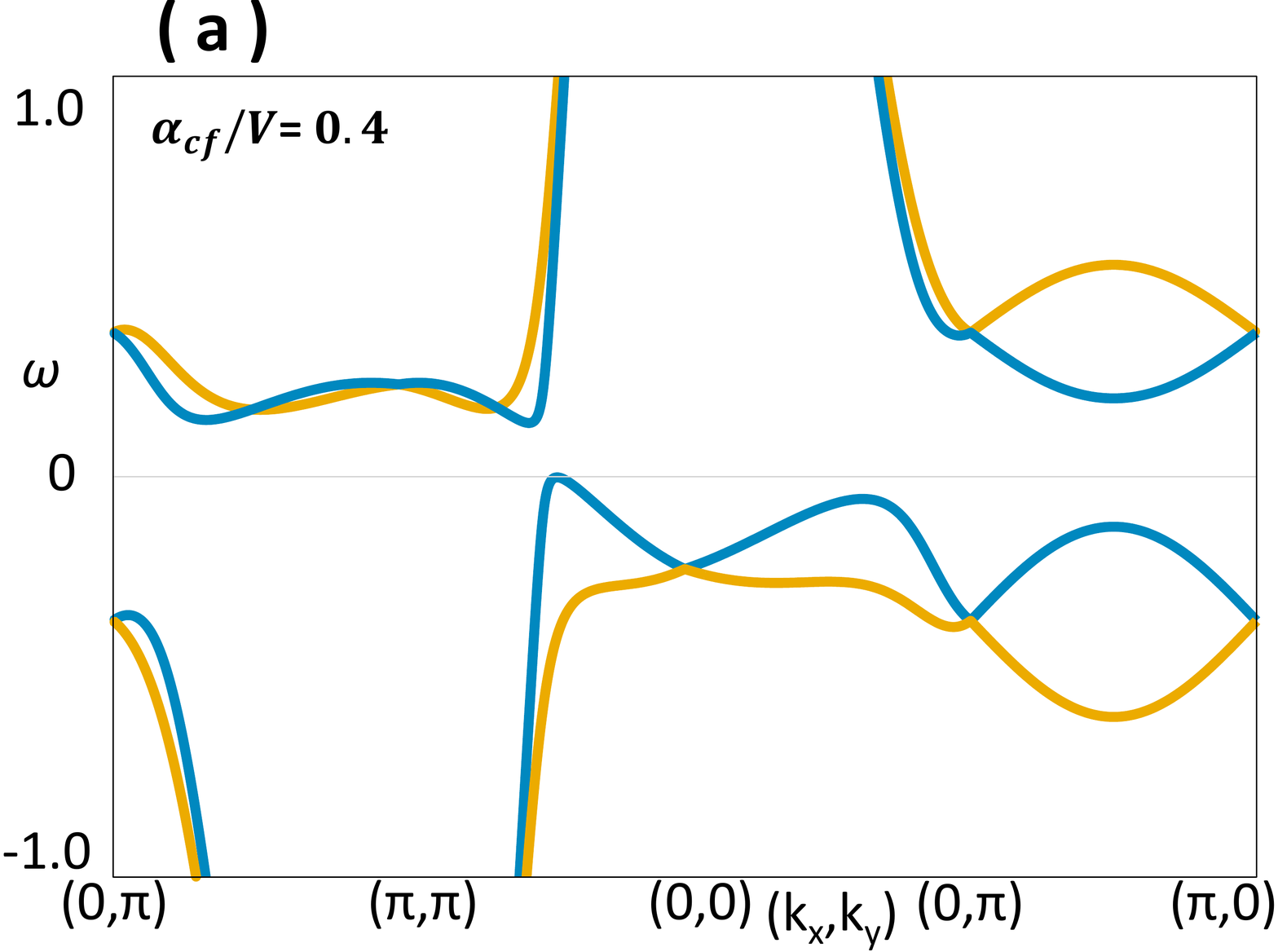}
\includegraphics[width=0.3\linewidth]{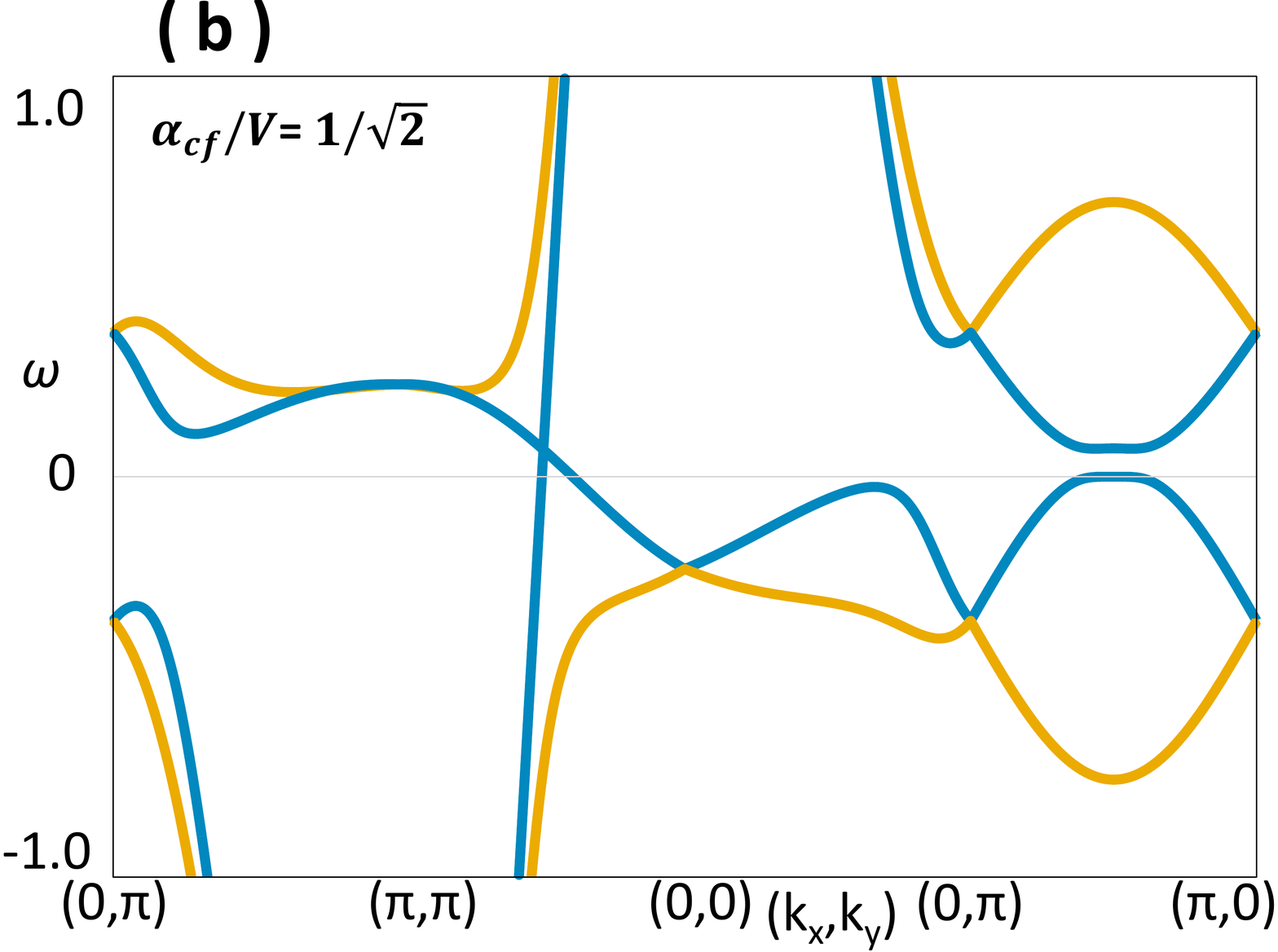}
\includegraphics[width=0.3\linewidth]{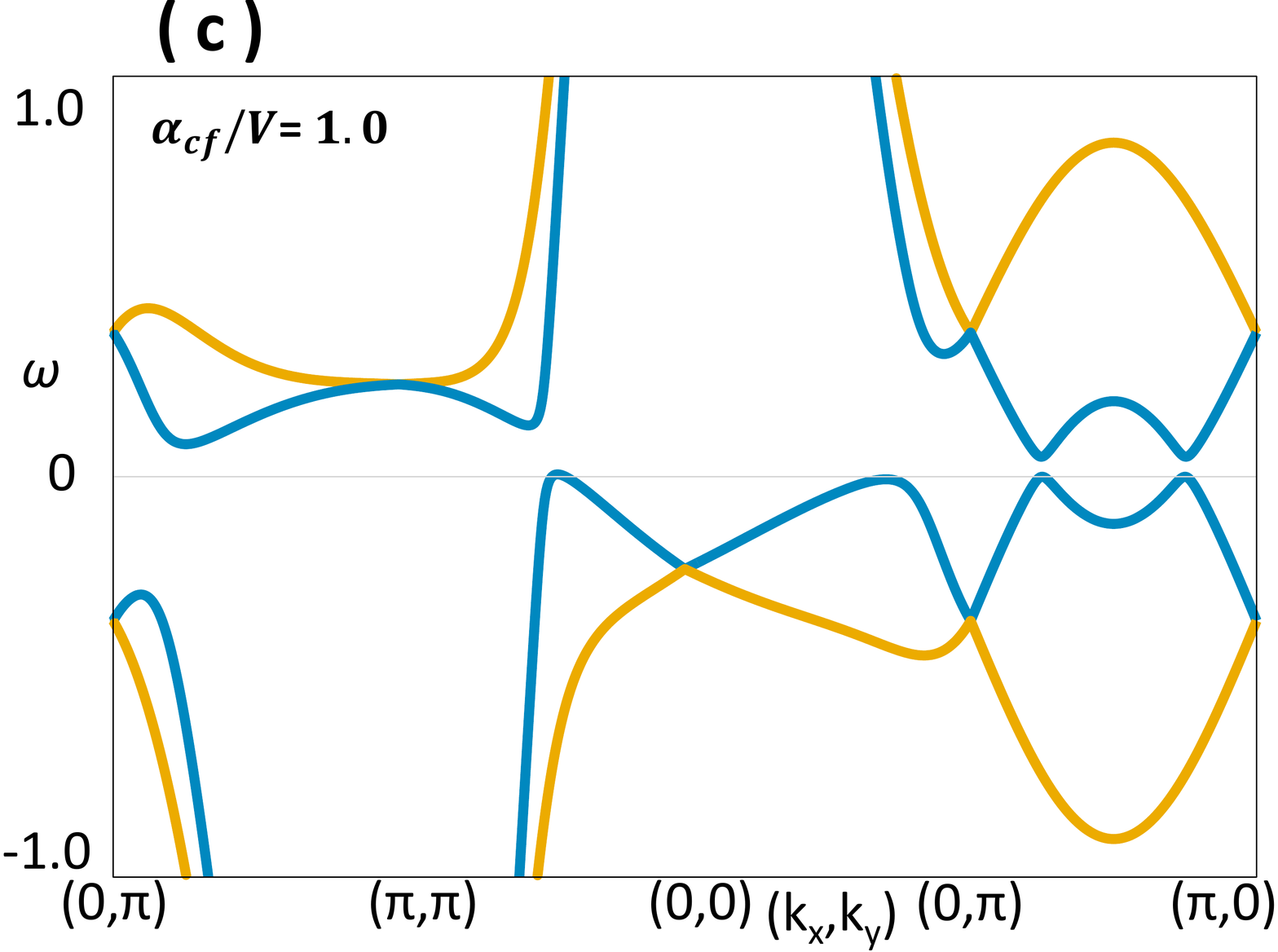}\\

\includegraphics[width=0.3\linewidth]{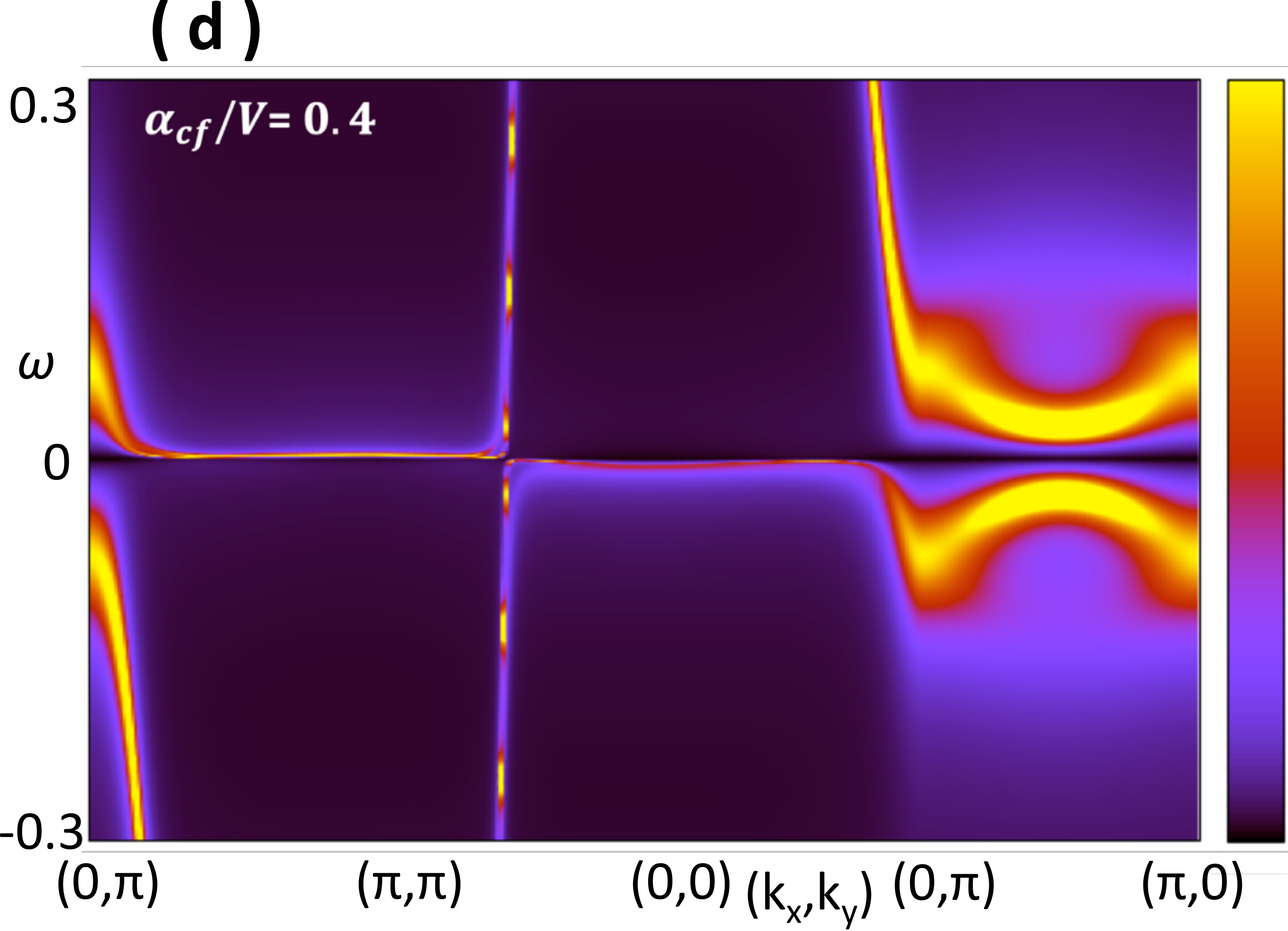}
\includegraphics[width=0.3\linewidth]{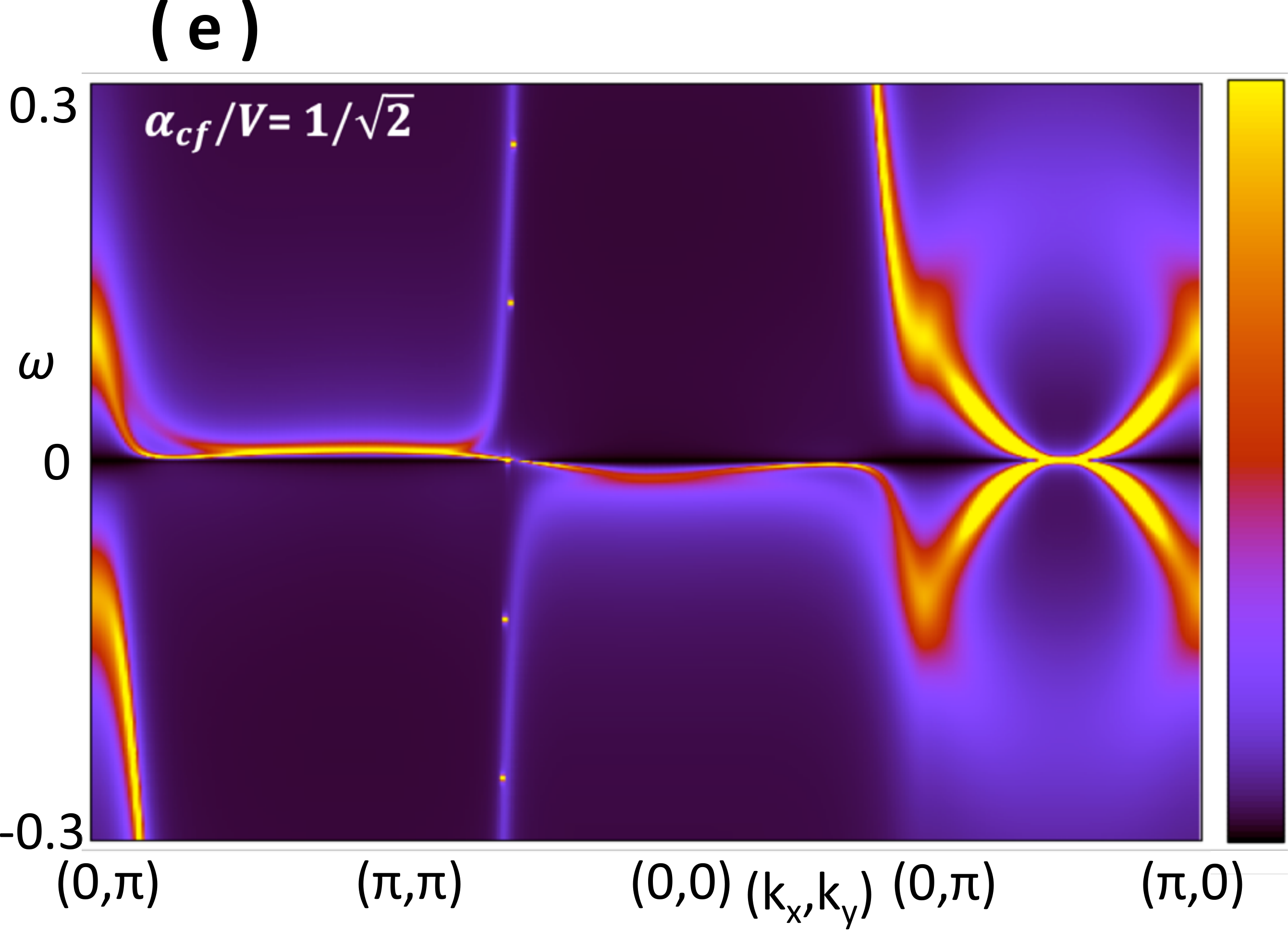}
\includegraphics[width=0.3\linewidth]{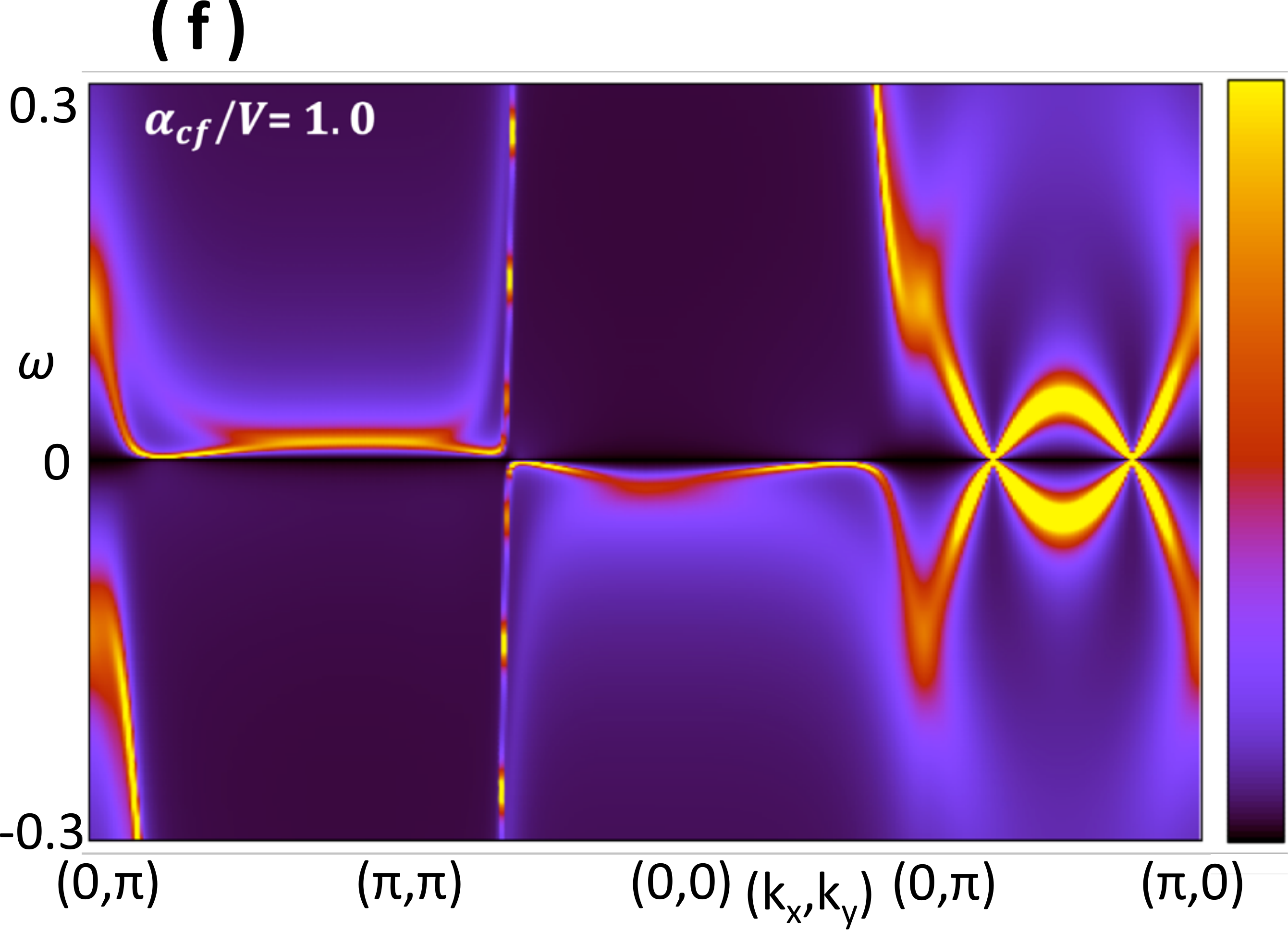}\\

\includegraphics[width=0.3\linewidth]{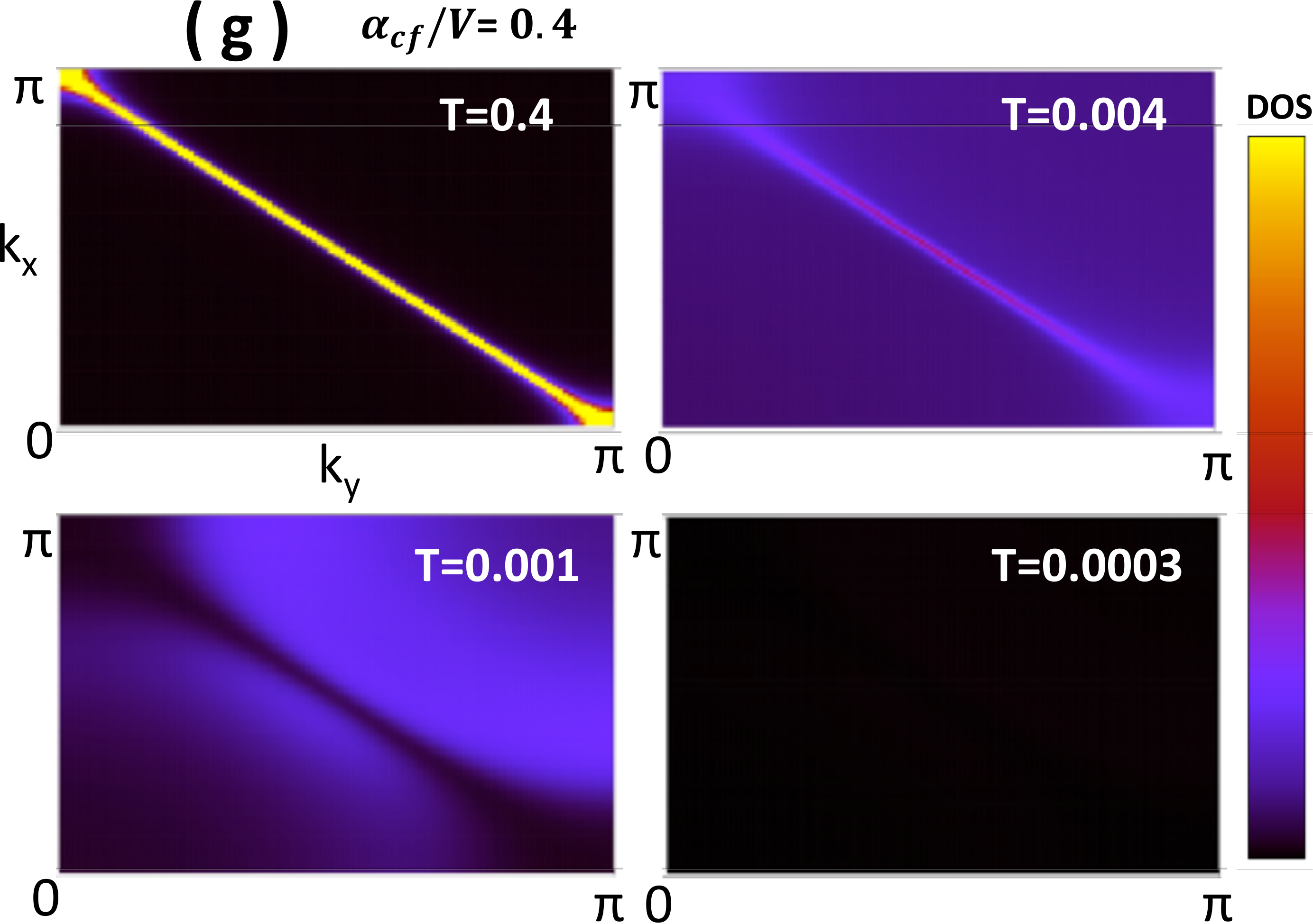}
\includegraphics[width=0.3\linewidth]{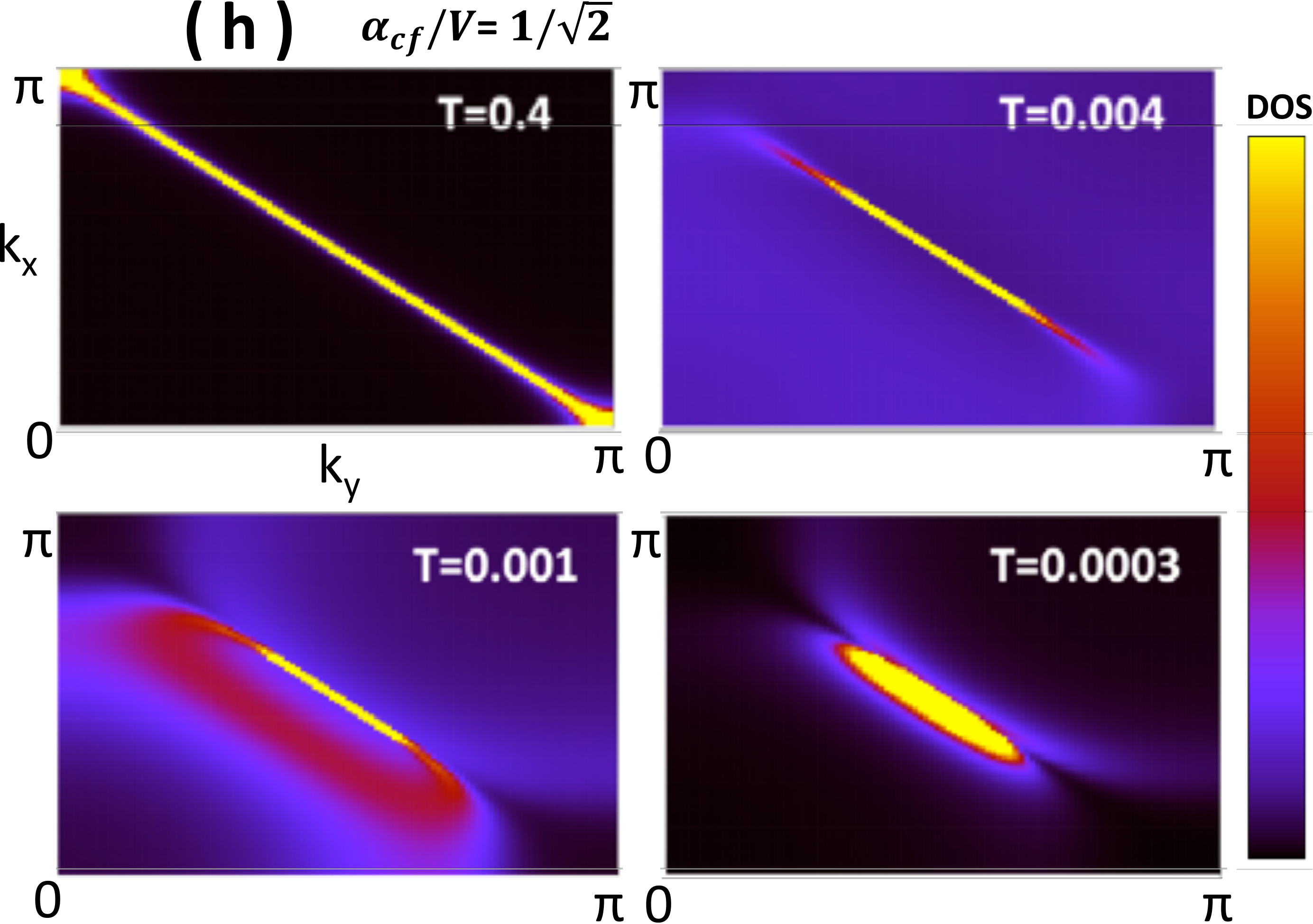}
\includegraphics[width=0.3\linewidth]{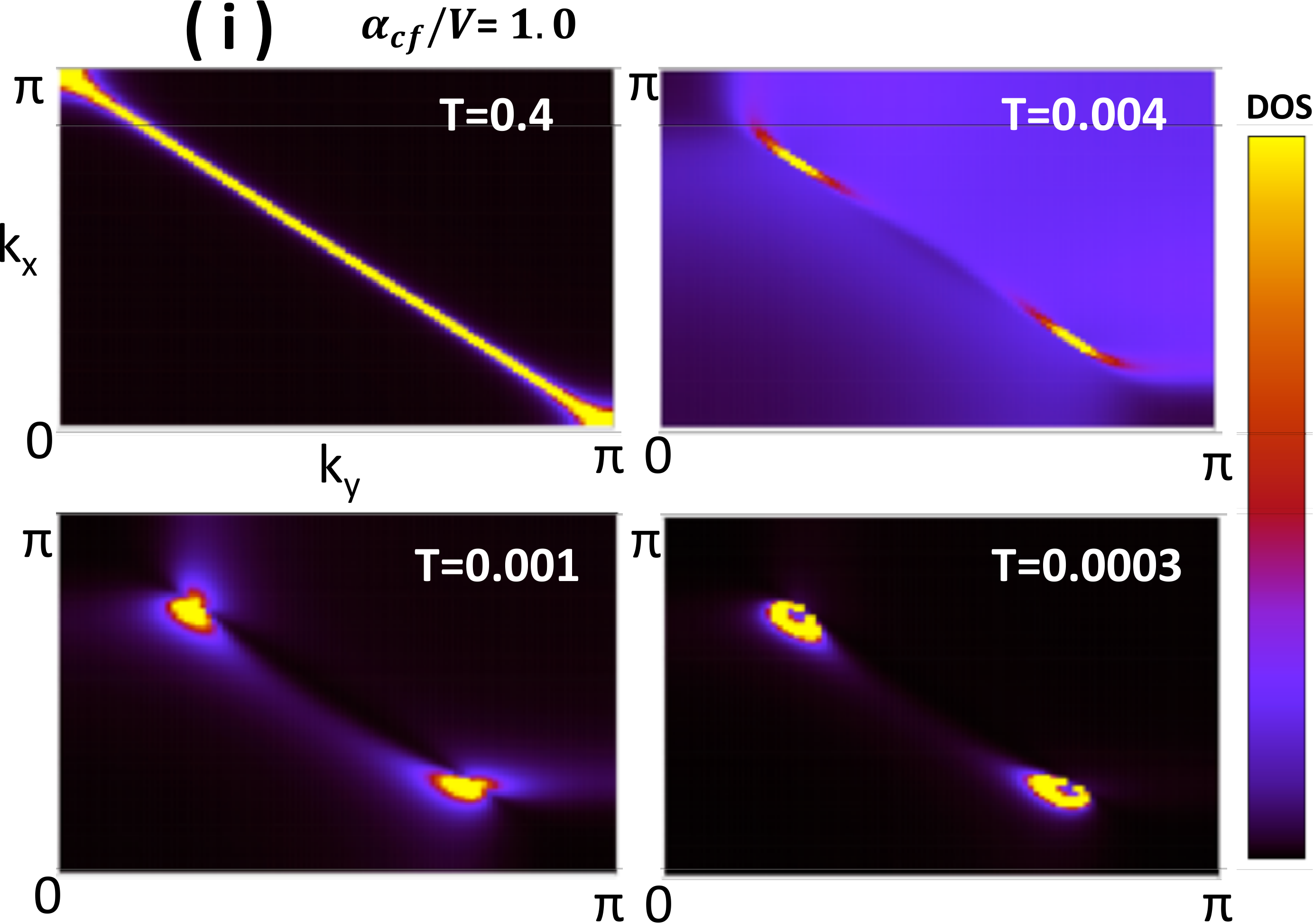}
\caption{ Panels (a)-(f) : Comparison between non-interacting and interacting band structures for different strengths of $\alpha_{cf}$. 
Panels (a)-(c) show non-interacting band structures. The colors in the non-interacting band structures (a)-(c) correspond to the different helical spin polarizations. 
Panels (d)-(f) show the interacting spectral functions at $T=0$.
Panels (g)-(i) show momentum resolved spectral functions at the Fermi energy for various temperatures. The parameters are $U$=2, $t_c$=1, $t_f$=-0.05, $\mu_c$=0, $\mu_f$=-1.0,$V$=0.36, $\alpha_{ff}$=0.05. 
\label{fig:Disp_example}}
\end{figure*}

After having studied the effect of the RSOC on the Kondo effect, we now study how the Kondo screening affects the band structure and particularly the Rashba splitting. In non- or weakly interacting metals including the RSOC, the band structure does not change when decreasing the temperature. On the other hand, the Kondo effect occurring in the $f$-electron material leads to a drastic change in the band structure at the Kondo temperature; $f$-electrons change from localized at high temperature to itinerant at low temperature. In Fig. \ref{fig:Disp_example}(a)-(f), we compare the band structure with and without Coulomb interaction at $T=0$ for different strengths of the RSOC.

To better understand the Rashba splitting, we block-diagonalize the non-interacting model and obtain the energy-momentum dispersions for two different helical bands, corresponding to the spin polarizations. The block-diagonalized Hamiltonians are
\begin{align}
\thickmuskip=0mu
\medmuskip=0mu
\thinmuskip=0mu
&{\cal H}=\nonumber\\
&(c^{\dagger}_{h\bm{k}} f^{\dagger}_{h\bm{k}})\left(
    \begin{array}{cc}
      \epsilon_c(\bm{k}) & V\mathalpha{+}h\alpha_{cf}(\bm{k}) \\
      V\mathalpha{+}h\alpha_{cf}(\bm{k}) & \epsilon_f(\bm{k})\mathalpha{+}h\alpha_{ff}(\bm{k}) \\
    \end{array}
  \right)\left(
    \begin{array}{cc}
    c_{h\bm{k}}\\
     f_{h\bm{k}}\\
     \end{array}
  \right)\nonumber\\
  \\
  &\alpha_{cf/ff}(\bm{k})=\alpha_{cf/ff} \sqrt{\sin^2{k_x}+\sin^2{k_y}}
  \end{align}
  which results in the eigenvalues  
  \begin{eqnarray}
E_{h,\pm}(\bm{k}) &=& \frac{1}{2}(\epsilon_c(\bm{k})\mathalpha{+}\epsilon_f(\bm{k})\mathalpha{+}h\alpha^{\prime}_{ff}(\bm{k}))\nonumber\\
   &&\pm \sqrt{\frac{1}{4}(\epsilon_c(\bm{k})\mathalpha{-}\epsilon_f(\bm{k})\mathalpha{-}h\alpha_{ff}(\bm{k}))^2\mathalpha{+}(V\mathalpha{+}h\alpha_{cf}(\bm{k}))^2}\nonumber\\\label{eq_helical_gap} 
\end{eqnarray}
where $h=\pm1$ is the helical index describing the spin polarization.
From these eigenvalues, we can derive the hybridization strength depending on the helical spin direction. 
 We see that the hybridization strength, and thus the hybridization gap, depends on the spin polarization  via $V+h\alpha_{cf}(\bm{k})$. 
 
 To demonstrate this effect, we show the non-interacting band structure for $\alpha_{cf}/V=0.4$, $\alpha_{cf}/V=0.71$, and $\alpha_{cf}/V=1.0$ 
 in Figs. \ref{fig:Disp_example}(a)-(c).  We have colored the helical spin direction $h=+1$ as orange and the $h=-1$ as blue.
Clearly visible in these non-interacting band structures is the arising spin-splitting due to the Rashba interaction corresponding to the difference between the $h=+1$ and $h=-1$ band.
The dependence of the hybridization strength on the spin polarization
is particularly visible  in Fig.  \ref{fig:Disp_example}(a) around $\bm{k}=(\pi/2, \pi/2)$. Around this momentum, we see that bands with helical polarization $h=-1$ (blue lines) open a small hybridization gap, while $h=+1$ (orange lines) open a large hybridization gap.
 For strong RSOC, $\alpha_{cf}>V/\sqrt 2\simeq 0.71 V$ shown in Figs. \ref{fig:Disp_example}(b)-(c), the gap in the $h=-1$ band is closed. 
 Exactly for $\alpha_{cf}=V/\sqrt 2$ and $\alpha_{ff}=0$, we find the gap closing at $\bm{k}=(\pi/2, \pi/2)$. 
For finite $\alpha_{ff}>0$, the gap closing occurs for large enough $\alpha_{cf}$, slightly shifted from $(\pi/2,\pi/2)$ in the Brillouin zone.

In the interacting system (Fig. \ref{fig:Disp_example}(d)-(f)), we see that the band structure is renormalized and smeared out away from the Fermi energy. 
Thus, even at $T=0$, the helical bands away from the Fermi energy do not form well-defined quasi-particles.  The gap closing, described for the non-interacting system, does also occur in the interacting system, as can be seen in Fig. \ref{fig:Disp_example}(e) and (f).

As explained above, in the interacting system, the band structure depends on the temperature. In Figs. \ref{fig:Disp_example}(g)-(i), we show the momentum resolved spectral functions at the Fermi energy, $\omega=0$, for different temperatures. At high temperatures above the Kondo temperature, the $f$-electrons are localized and thus absent from the Fermi surface. We see that, independent of the strength of the Rashba interaction,  the Fermi surface consists of a single unsplit band at $T=0.4$.
This band corresponds to the $c$-electron band. Lowering the temperature to $T=0.004$ $\sim$ $T=0.001$, the $f$-electrons start to become itinerant; the Fermi surface begins to change. Particularly for strong RSOC, we see that a ring-like band structure develops at the Fermi surface, which is a manifestation of the Rashba splitting in the band structure. 
At temperatures below the Kondo temperature, the $f$-electrons are itinerant and hybridize via the RSOC with the $c$-electrons. Because of the hybridization between $c$- and $f$-electrons, a gap opens for small RSOC, $\alpha_{cf}/V=0.4$, and the system becomes insulating; the Fermi surface vanishes in Fig. \ref{fig:Disp_example}(g) at $T=0.0003$.
For large RSOC, $\alpha_{cf}>V/\sqrt{2}$, the gap closes in the $h=-1$ band.
Thus, a spin-polarized Fermi surface due to the $h=-1$ band is observed at low temperatures for $\alpha_{cf}>V/\sqrt{2}$, see Fig. \ref{fig:Disp_example}(h)-(i) at $T=0.0003$.

\begin{figure}
\includegraphics[
height=1.2in,
width=2.8in
]{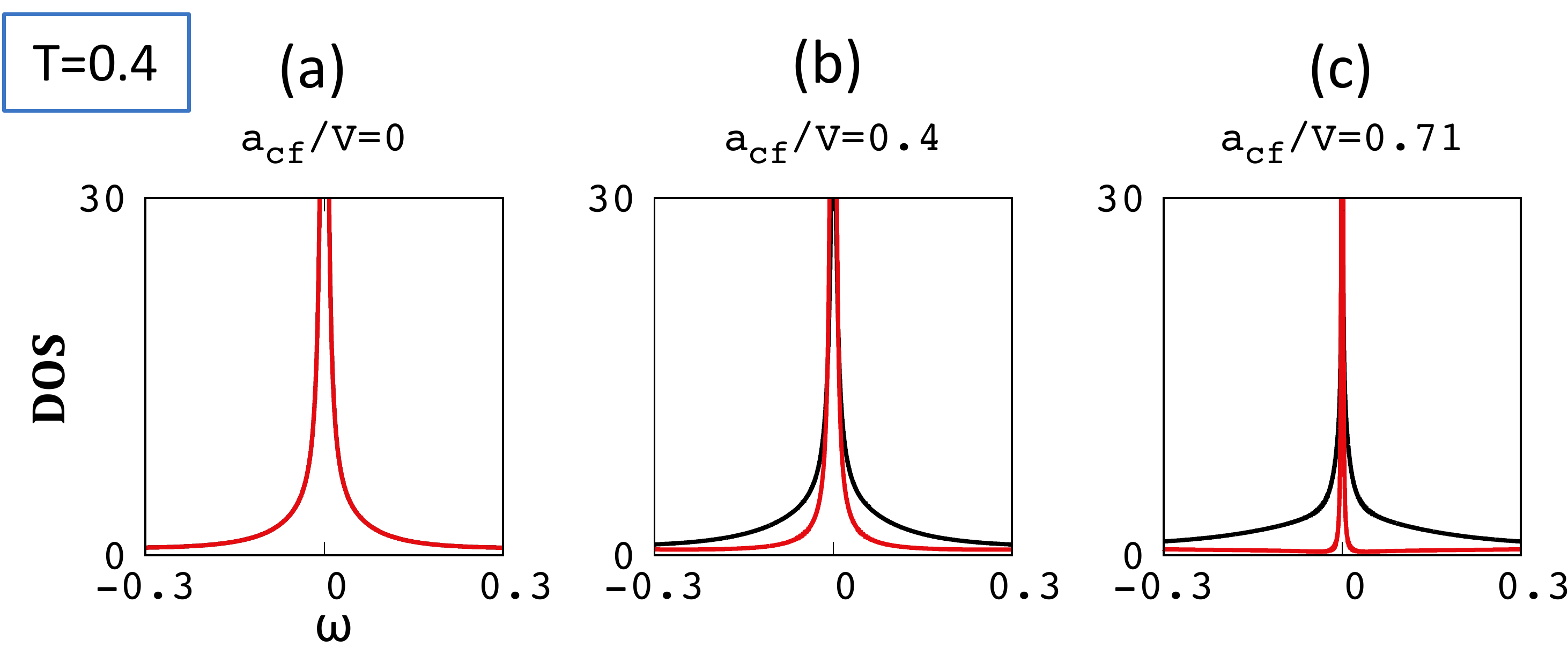}
\includegraphics[
height=1.2in,
width=2.8in
]{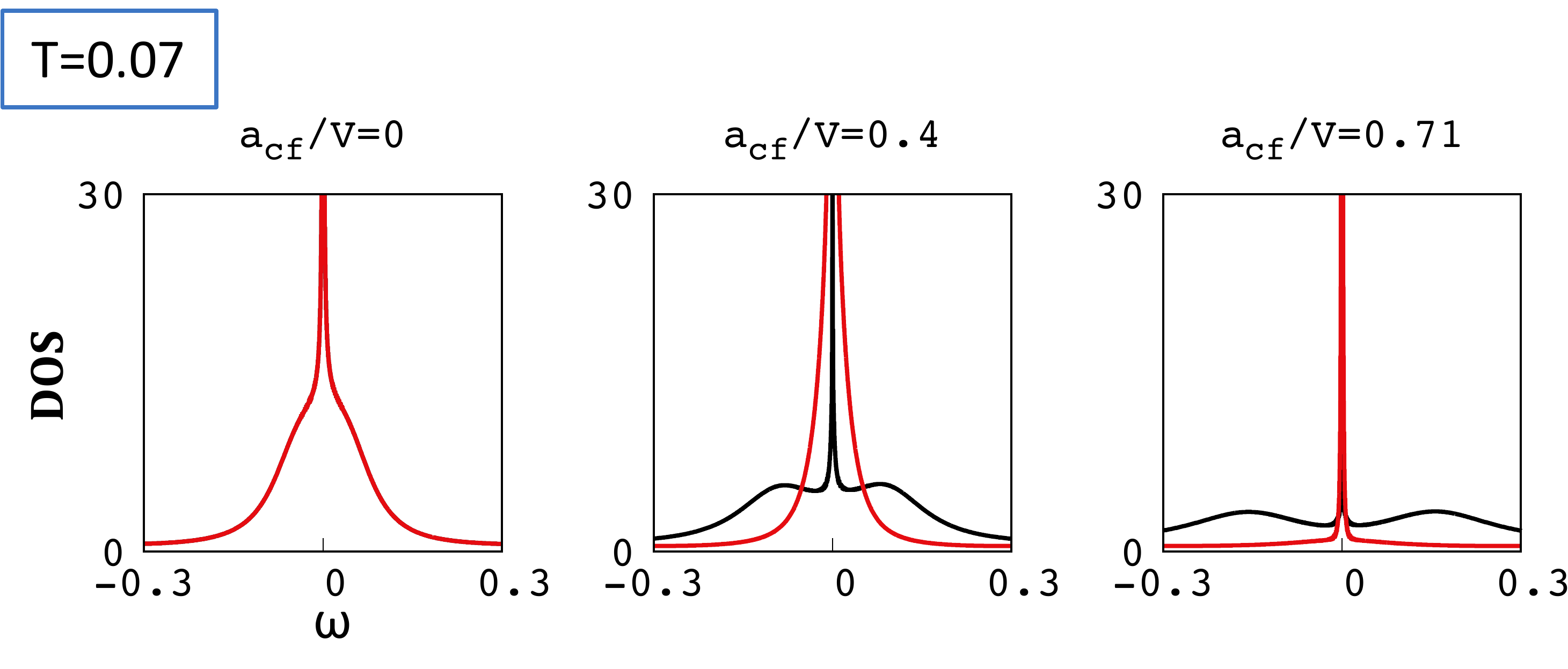}
\includegraphics[
height=1.2in,
width=2.8in
]{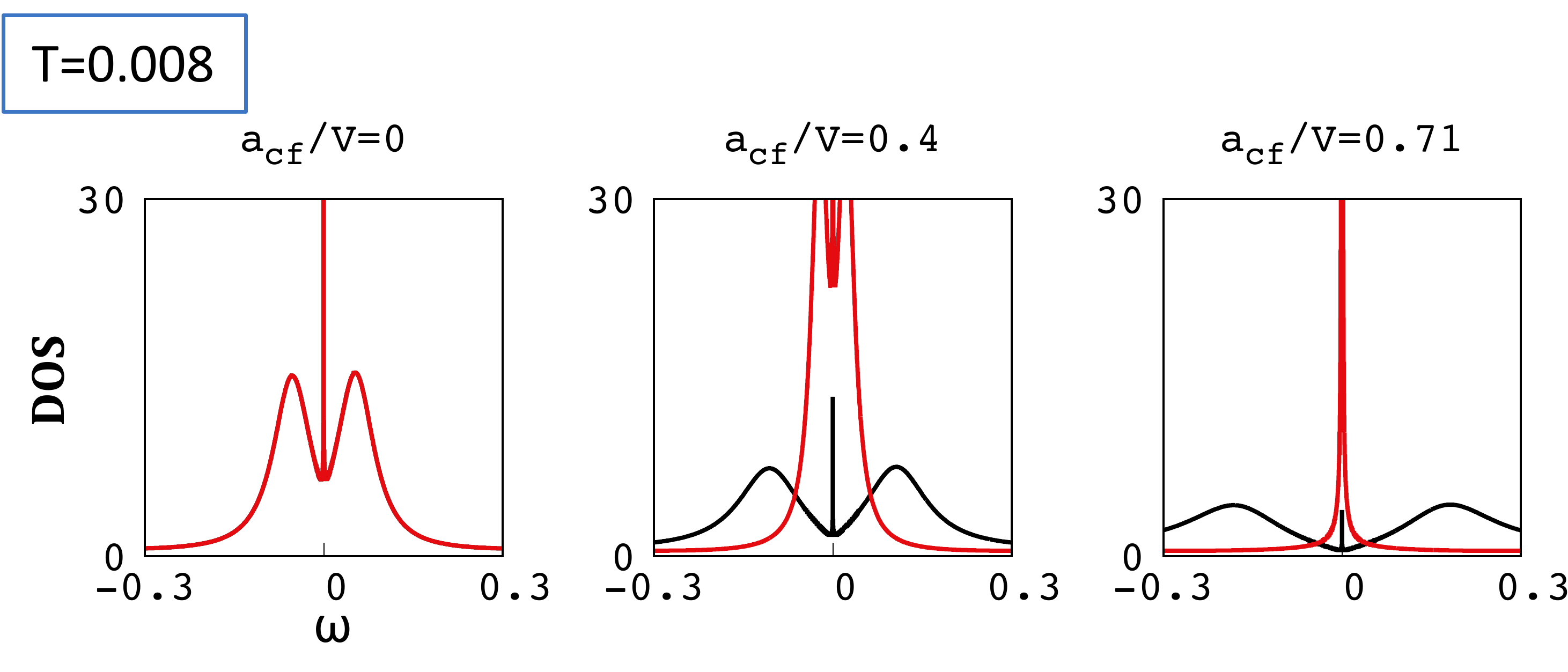}
\includegraphics[
height=1.2in,
width=2.8in
]{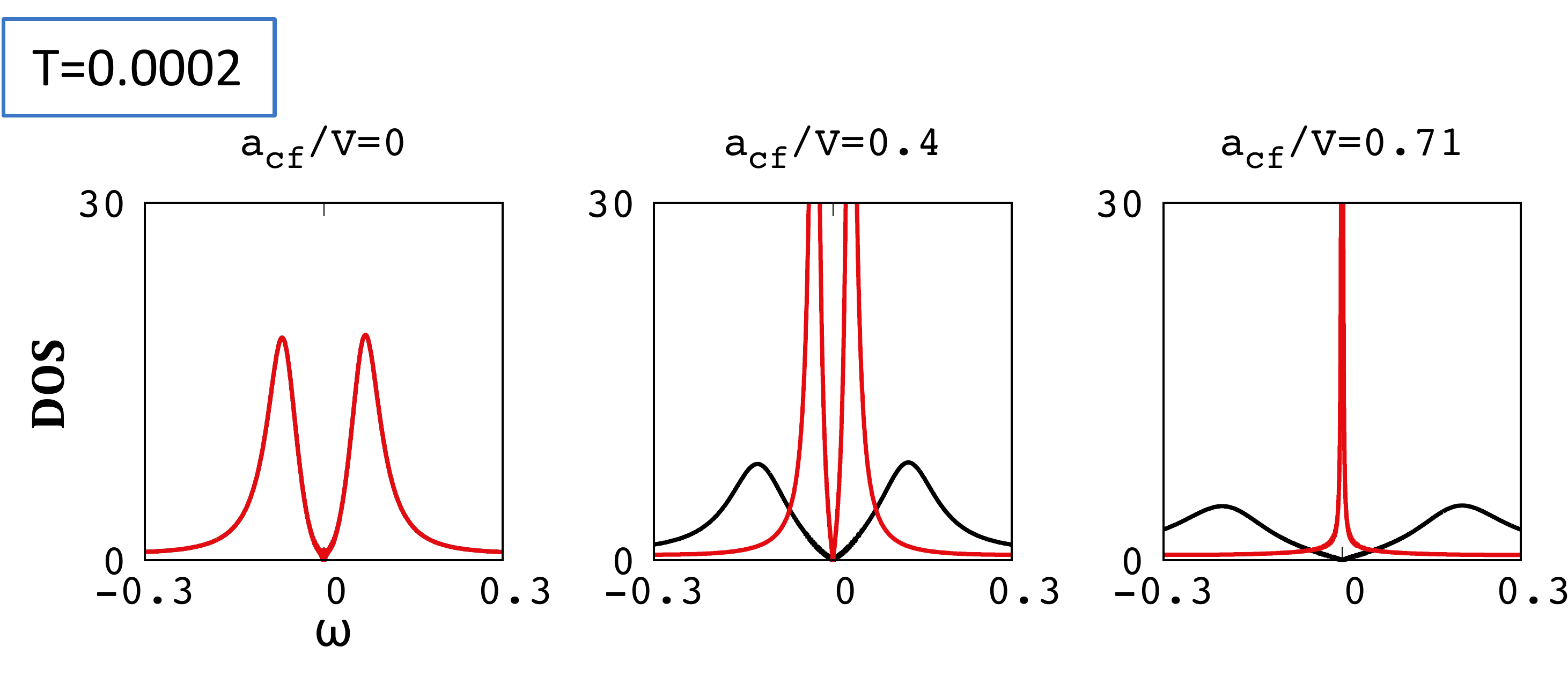}
\caption{Spectral functions for the $h=+1$ band (black lines) and the $h=-1$ band (red line) at $\bm{k}=(\pi/2,\pi/2)$ for $\alpha_{cf}/V=0$ (left panels),  $\alpha_{cf}/V=0.4$  (middle panels), and   $\alpha_{cf}/V=0.71$  (right panels)
Other parameters are $U=2$, $t_c=1$, $t_f=-0.05$, $\mu_c=0$, $\mu_f=-1.0$, $V=-0.36$, $\alpha_{ff}=0$.}
\label{fig:RS}
\end{figure}
\begin{figure}
\includegraphics[
height=1.2in,
width=2.8in
]{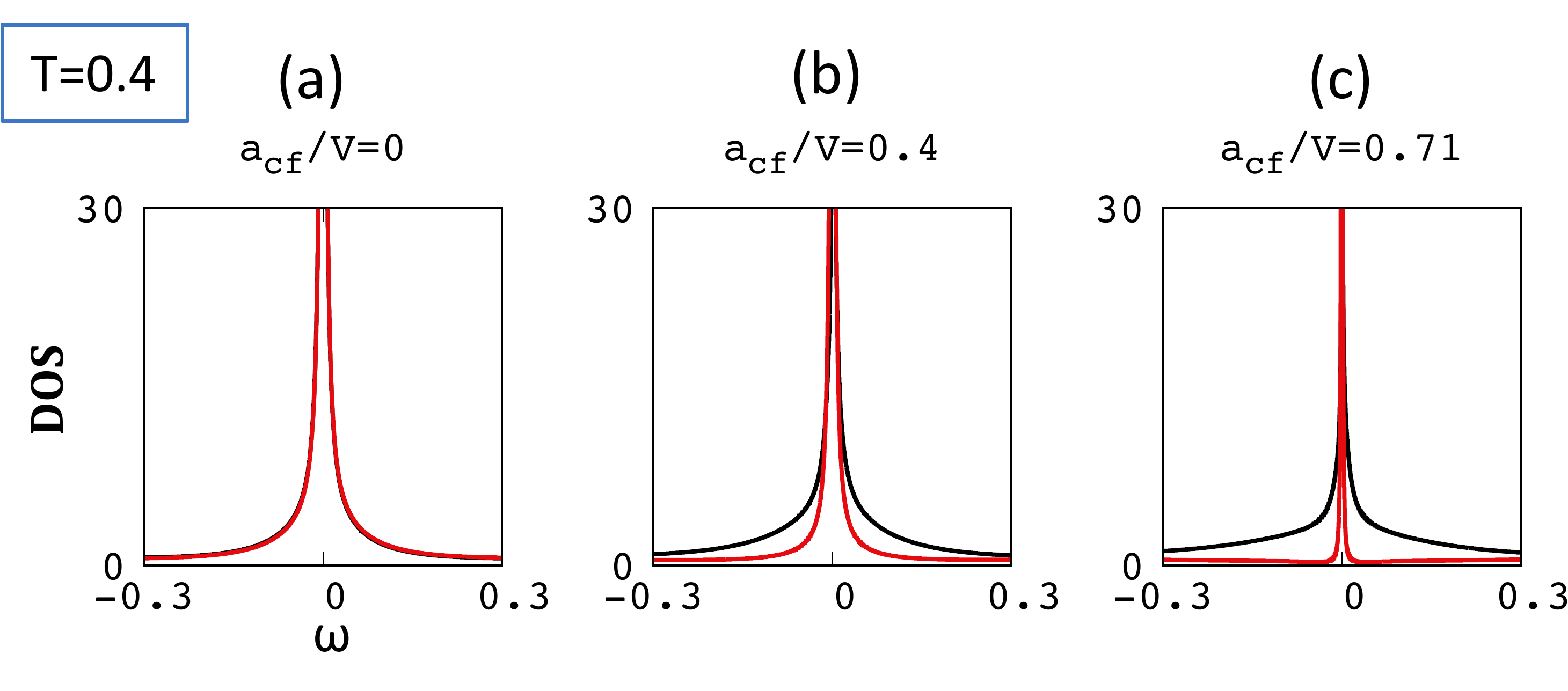}
\includegraphics[
height=1.2in,
width=2.8in
]{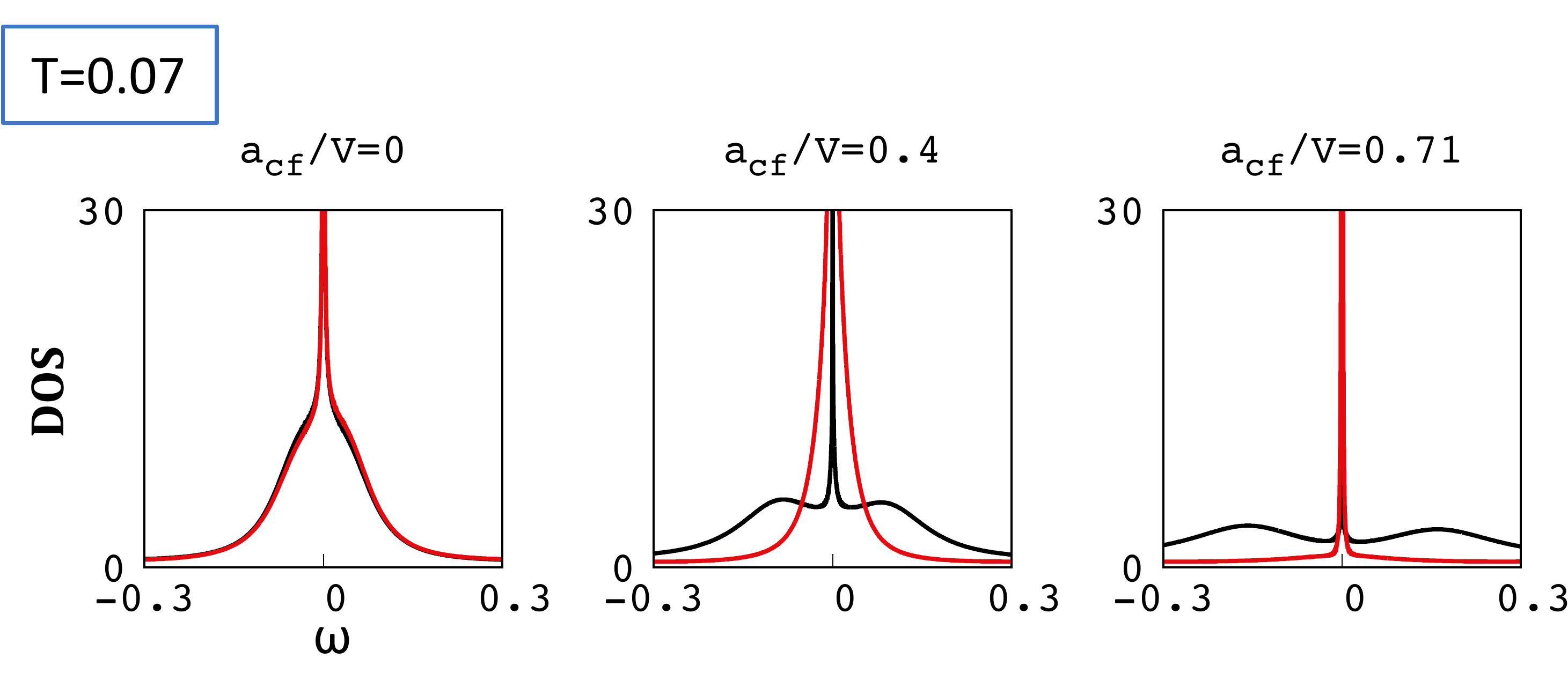}
\includegraphics[
height=1.2in,
width=2.8in
]{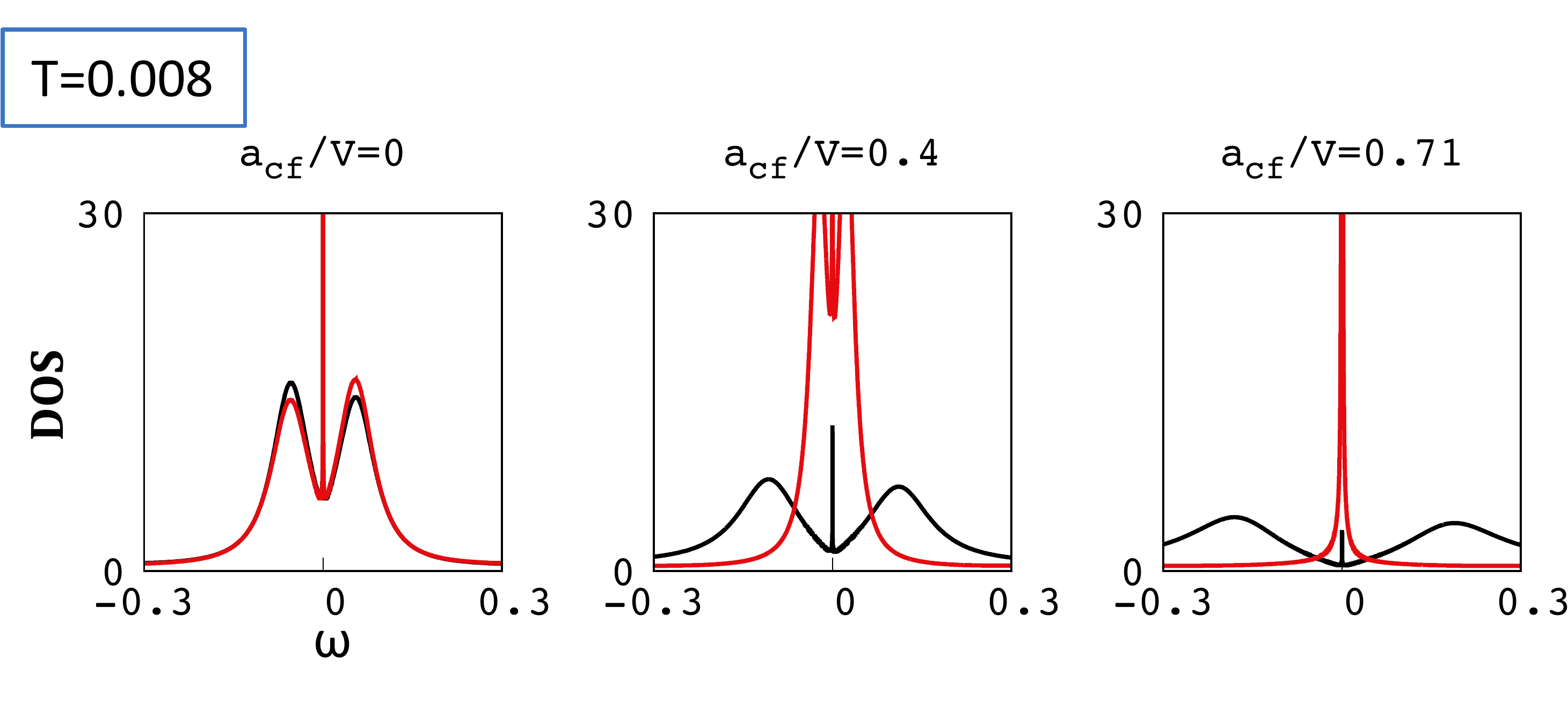}
\includegraphics[
height=1.2in,
width=2.8in
]{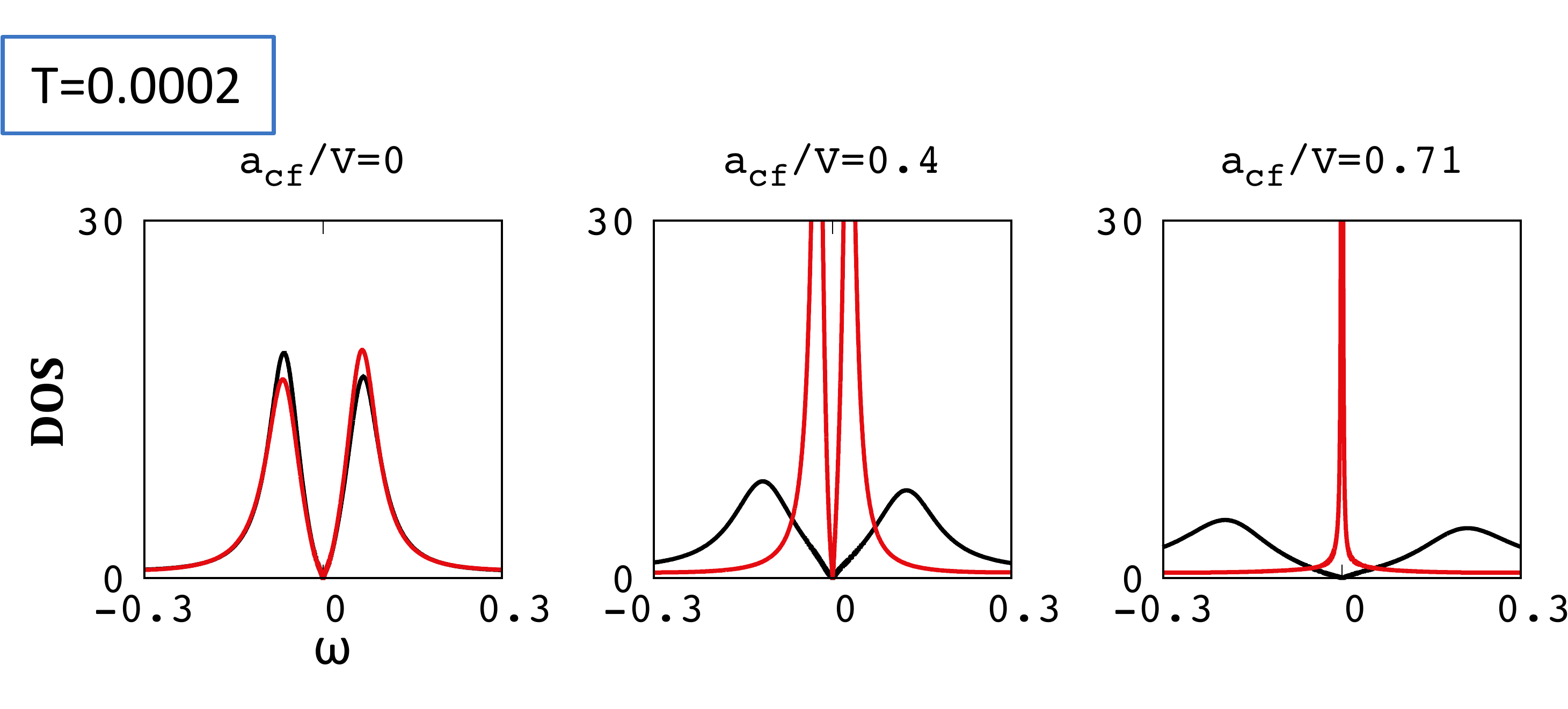}
\caption{Spectral functions for the $h=+1$ band (black lines) and the $h=-1$ band (red line) at $\bm{k}=(\pi/2,\pi/2)$ for $\alpha_{ff}=0.05$.  The rest of the parameters are the same as in Fig. \ref{fig:RS}.}
\label{fig:RS1}
\end{figure}

To obtain some more information about the temperature dependent Rashba splitting, we show spectral functions of the  $h=+1$ and the $h=-1$ band at momentum $\bm{k}=(\pi/2, \pi/2)$ for different temperatures in Fig. \ref{fig:RS} ($\alpha_{ff}=0$) and Fig. \ref{fig:RS1} ($\alpha_{ff}=0.05$). We show spectral functions for  $\alpha_{cf}/V=0$ (a), $\alpha_{cf}/V=0.4$ (b), and $\alpha_{cf}/V=0.71$ (c) and note again that the system changes for $\alpha_{cf}/V=0.71$ from an insulator to a metal at low temperature.

The top panels show the spectral functions at high temperature. The  $f$-electrons are absent from the Fermi energy at this temperature due to the strong Coulomb repulsion. As a consequence, we only find the peaks of the conduction electrons for both helical spin directions, located at the Fermi energy for $\bm{k}=(\pi/2, \pi/2)$. 
Because there is no direct RSOC in the conduction electron band, these peaks are unsplit; the peak of $h=+1$ band lies at the same energy as the peak of the $h=-1$ band. 
However, the height and the width of the peaks in the $h=+1$  and the $h=-1$ bands are different, if $\alpha_{cf}\neq 0$. 
This phenomenon arises due to the self energy of the $f$-electrons.
The self energy of the $f$-electrons, which exhibits a strong peak in the imaginary part at high temperatures, shifts the spectral weight of the $f$-electrons away from the Fermi energy. Thus, $f$-electrons are localized. $c$-electrons have a peak at the Fermi energy at high temperature, because they cannot form singlets with the $f$-electrons. Nevertheless, $c$-electrons can undergo  scattering processes with the $f$-electrons, which depend on the strength of hybridization. This process gives $c$-electrons an effective finite life-time which broadens the $c$-electron peak at the Fermi energy. In this system, the strength of hybridization depends on $h=\pm1$ and therefore the height and the width of the $c$-electron peak depends on the helical spin polarization.
This is a remarkable effect, as it leads to an asymmetry between $h=+1$ and $h=-1$ bands at high temperature, possible even room temperature, in the conduction electron band. Such an asymmetry might be used for spintronic devices as it can induce an electromagnetic effect\cite{PhysRevB.97.115128}.

This asymmetry between helical bands does not only lead to different high temperature spectral functions, but has also a distinctive effect when decreasing the temperature. 
While for $\alpha_{cf}=0$ the spectral functions are degenerate for all temperatures, the temperature dependence of the correlation effects in the spectral functions depend on the helical spin polarization  for finite RSOC, $\alpha_{cf}>0$.
 For  $\alpha_{cf}/V>0$, the $h=+1$ band, which has a large hybridization strength and thus involves strong scattering from the $f$-electrons, changes to a triple peak structure at $T=0.07$. Spectral weight is transferred away from the Fermi energy. On the other hand, the $h=-1$ band remains nearly unchanged.
Only when lowering the temperature to $T=0.008$, correlation effects are  visible in both helical bands for $\alpha_{cf}/V=0.4$. While the $f$-electrons become itinerant, the $f$- and the $c$-electrons hybridize and thus the spectral weight at the Fermi energy is reduced. For $\alpha_{cf}/V=0.4$, the spectral weight at the Fermi energy completely vanishes at $T=0.0002$. On the other hand, for $\alpha_{cf}/V=0.71$ the effective hybridization $V+h\alpha_{cf}$ vanishes for the $h=-1$ band at $\bm{k}=(\pi/2, \pi/2)$. As a result, the peak at the Fermi energy at high temperature is very narrow for this band. Furthermore, when decreasing the temperature, this peak in the $h=-1$ band does not show any correlation effects and persists  to $T=0$. 

Including the RSOC directly into the $f$-electron band ($\alpha_{ff}>0$), shown in Fig. \ref{fig:RS1}, we observe that the above described physics persists. An additional feature is that the spectral weight is asymmetrically transferred away from the Fermi energy. For momentum $\bm{k}=(\pi/2, \pi/2)$, the spectral weight of the $h=+1$ band is transferred to negative energies, while the $h=-1$ band is transferred to positive energies.

\begin{figure}[t]
\includegraphics[
height=2.3in,
width=3.0in
]{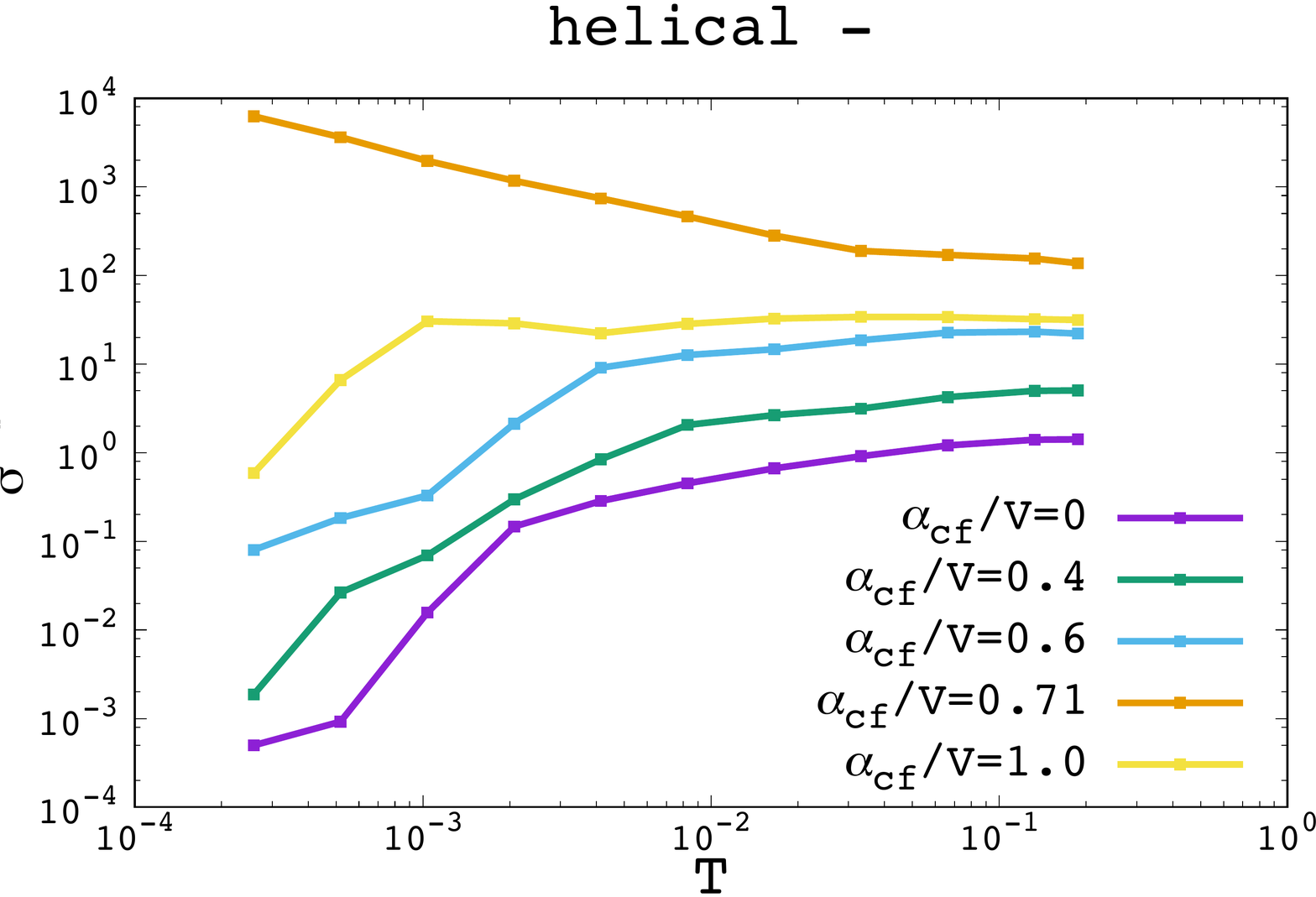}
\includegraphics[
height=2.3in,
width=3.0in
]{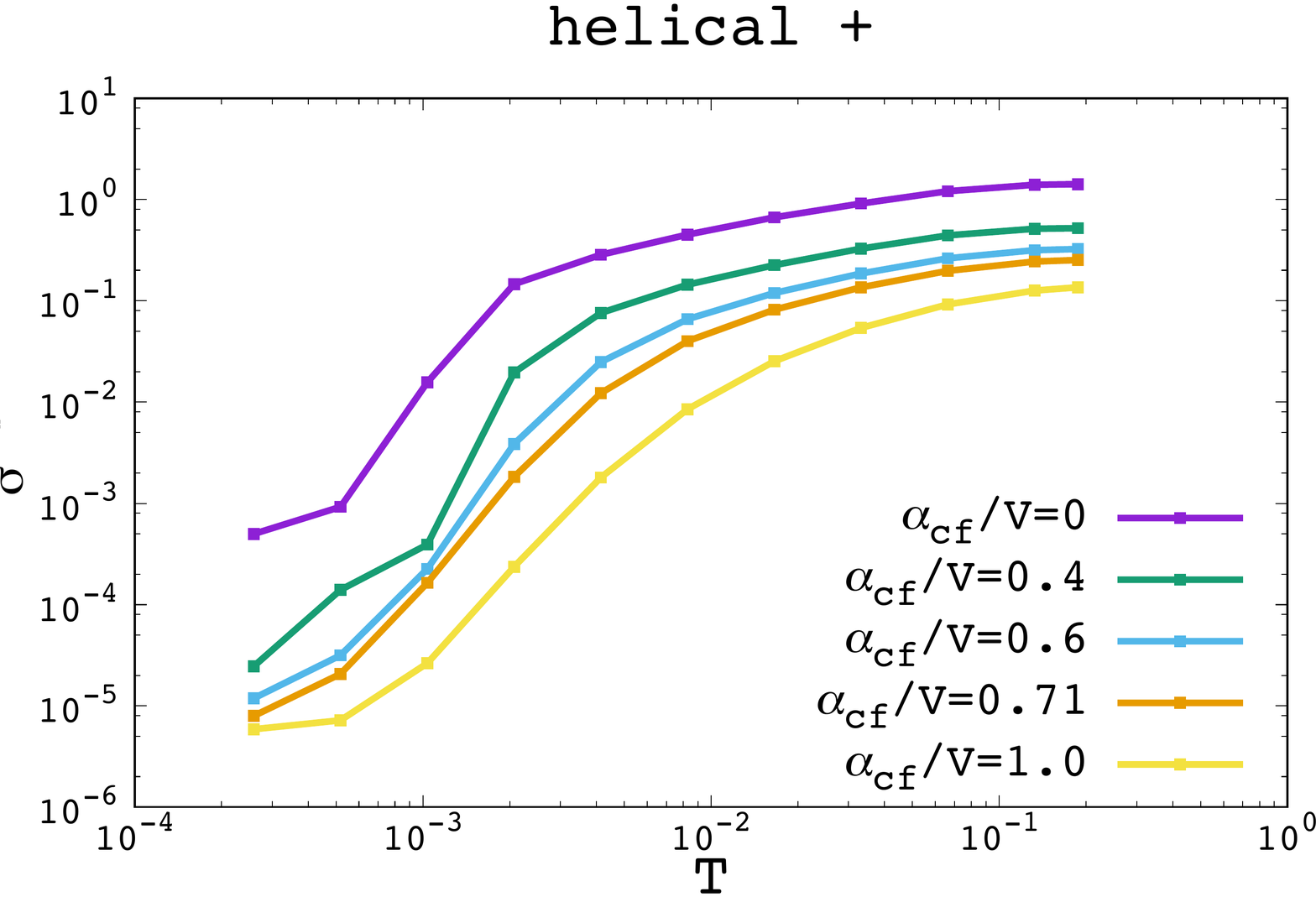}
\caption{Conductivity for the $h=-1$ (top panel) and the $h=+1$ band (bottom panel). The parameters are $U=2$, $t_c=1$, $t_f=-0.05$, $\mu_c=0$, $\mu_f=-1.0$,$V=-0.36$, $\alpha_{ff}=0$.}
\label{fig:Cond_ff0}
\end{figure}
The above-described difference of the effective hybridization and the possible gap closing is not only visible in the spectral functions, but has also a distinct impact on transport properties.
We thus calculate the conductivity depending on the helical spin polarization using DMFT/NRG.
This helical conductivity is calculated via the Kubo-formula
\begin{eqnarray}
&\sigma^{h}_{xx}=C \int d\omega \frac{\partial f(\omega)}{\partial \omega} \sum_{\bm{k}} \frac{\partial {\cal H}_h}{\partial k_x} A_h(\bm{k},\omega) \frac{\partial {\cal H}_h}{\partial k_x} A_h(\bm{k},\omega)\nonumber\\
\\
&{\cal H}_h=\frac{1}{\sqrt{2}}\left(
    \begin{array}{cc}
      \epsilon_c(\bm{k}) & V_h \\
      V_h & \ \epsilon_f(\bm{k})+\mu_f+\Sigma+h\alpha_{ff}(\bm{k}) \\
      \end{array}
  \right),
\end{eqnarray}
where $A_h$ is the  spectral function of the helical band with $h=\pm 1$ and $f(\omega)$ is the Fermi function at temperature T. $V_h$ is the effective hybridization, $V+h\alpha_{cf}(\bm{k})$. In this study, we neglect vertex corrections, which vanish in DMFT. We note, however, that vertex corrections usually do not change the conductivity qualitatively\cite{PhysRevB.94.079907}.

Fig. \ref{fig:Cond_ff0}  shows the temperature dependence of the conductivity on the helical spin polarization for  different $\alpha_{cf}$ and $\alpha_{ff}=0$. We have not observed a noteworthy difference in the conductivity when including $\alpha_{ff}=0.05$. 
The conductivity for $\alpha_{cf}=0$, purple lines, does not depend on the helical polarization.  Increasing the RSOC,  we see that the conductivity of the $h=-1$ band is increased, while the conductivity of the $h=+1$ band is decreased. 
Even at high temperatures, the conductivity of each helical band is different, because the strength of the scattering between $c$-electrons and localized $f$-electrons depends on the helical spin direction. Because of these different scattering strengths, the peak at high temperatures in the $h=-1$ band shown in Fig. \ref{fig:RS} is higher than in the $h=+1$ band. Thus, the conductivity for the $h=-1$ band is larger even at high temperatures. 
For $\alpha_{cf}/V<0.71$, the system forms an insulator at $T=0$, which is confirmed by the vanishing conductivity at low temperatures.
On the other hand, for $\alpha_{cf}/V=0.71$, the gap closes and we see that the conductivity of the $h=-1$ band diverges for $T\rightarrow 0$. Increasing further $\alpha_{cf}$, we observe that the conductivity slightly decreases. Although the system has a Fermi surface for $\alpha_{cf}/V>0.71$, the band structure changes. While exactly at $\alpha_{cf}/V=1/\sqrt{2}$, the bands touch the Fermi energy with a quadratic dispersion, for larger RSOC the dispersion changes to linear, see Fig. \ref{fig:Disp_example} (e)-(f). This change of the band structure around the Fermi energy results in a decrease of the conductivity.
While the system is a normal metal at high temperature, we see that for $\alpha_{cf}/V>1/\sqrt{2}$ it forms a helical half-metal below the Kondo temperature, where the electrical current is completely carried by the electron band with helical $h=-1$ spin polarization.

\section{Conclusion}
In summary, we have comprehensively studied  the interplay between the RSOC and the Kondo effect in noncentrosymmetric $f$-electron materials, particularly investigating a Rashba-like hybridization, which can be expected to occur in these materials.  
We have shown that the Kondo coupling  becomes spin anisotropic due to the presence of the RSOC. However, scaling theory shows that when lowering the temperature towards the Kondo temperature, a SU(2) symmetric Kondo singlet is formed. Generally, we see that the RSOC suppresses the Kondo temperature in noncentrosymmetric $f$-electron systems. We think that this suppression of the Kondo temperature can be experimentally observed by comparing different $f$-electron superlattices where the strength of the RSOC depends on the lattice structure.

Furthermore, we have demonstrated that the Kondo effect has also impact on the Rashba splitting observed in the band structure. While in non-interacting systems, the band structure does not change when decreasing the temperature, in noncentrosymmetric $f$-electron materials, the $f$-electrons change from localized at high temperatures to itinerant at low temperatures. Due to the localization of the $f$-electrons at high temperatures, the $c$-electrons form a Fermi surface at high temperatures corresponding to their band structure without hybridization. Remarkable, the width and the height of the $c$-electron peaks at the Fermi surface depend on the helical spin polarization, because the effective hybridization between the $c$- and the $f$-electrons depends on the spin polarization. 
These different hybridization strengths can also be observed in the changes occurring due to the Kondo effect when lowering the temperature. Finally, we have shown  that for strong  RSOC, the effective hybridization of the $h=-1$ helical polarization disappears at certain points in the Brillouin zone and the Kondo insulator changes into a helical half-metal where only the helical $h=-1$ band is present at the Fermi energy. 
In such a helical half-metal, both charge and spin currents can flow simultaneously.

\begin{acknowledgments}
We thank Y.Yanase and T. Yoshida for valuable advice and discussion. This work is partly supported by JSPS KAKENHI Grants No. JP18H04316, and No. JP18K0351.
Computer simulations have been performed on the supercomputer of the ISSP in the University of Tokyo.
\end{acknowledgments}

\appendix
\section{Deriving the Kondo lattice model\label{APP_KLM}}
We derive the Kondo lattice Hamiltonian from the periodic Anderson model, Eq.(\ref{X}) in the main text, setting $\mu_f=-U/2$ and assuming $U \gg V,\alpha_{cf}, \alpha_{ff}, t_f$. These parameters result in a half-filled lattice. Thus, a single $f$-electron occupies every site, which we denote as $\ket{1}$. We treat $V,\alpha_{cf}$ as perturbation which mixes the ground state $\Ket{1}$ with excited states $\Ket{i}$ or $\Ket{\bar{i}}$, where $\Ket{i}$ ($\Ket{\bar{i}}$) corresponds to the state in which the $f$-orbital at site $i$ is filled with two (no) electrons.  Then the effective interaction term derived by  second order perturbation theory is
\begin{alignat}{1}
\thickmuskip=0mu
\medmuskip=0mu
\thinmuskip=0mu
&{\cal H}_{mix}\mathalpha{=}\sum_{\bm{k}}(c^{\dagger}_{\bm{k}\alpha}f_{\bm{k}\beta}\mathalpha{+} h.c.)\biggl[V\delta_{\alpha\beta}\mathalpha{+}\alpha_{cf}(\sigma^x\sin{k_y}\mathalpha{-}\sigma^y\sin{k_x})_{\alpha\beta}\biggr]\nonumber\\
\\
&{\cal H}^{int}_{eff} =\sum_i \bigl[ \Bra{1}{\cal H}_{mix} \Ket{i} \Braket{i|\frac{1}{E_{11}-{\cal H}}|i} \Bra{i} {\cal H}_{mix} \Ket{1}\nonumber\\
& \ \ \ \ \ \ \ \ \ \ +\Bra{1}{\cal H}_{mix} \Ket{\bar{i}} \Braket{\bar{i}|\frac{1}{E_{11}-{\cal H}}|\bar{i}} \Bra{\bar{i}} {\cal H}_{mix} \Ket{1}\bigr] \nonumber\\
&=\sum_i \sum_{\bm{k},\bm{k^{\prime}}} \biggl[c^{\dagger}_{\bm{k^{\prime}},\alpha} (f_{\bm{k^{\prime}},\delta})_i \frac{1}{E_{11}-{\cal H}_{ii}} (f^{\dagger}_{\bm{k},\gamma})_i c_{\bm{k},\beta}\nonumber\\
& \ \ \ \ + (f^{\dagger}_{\bm{k},\gamma})_i c_{\bm{k},\beta} \frac{1}{E_{11}-{\cal H}_{\bar{i}\bar{i}}} c^{\dagger}_{\bm{k^{\prime}},\alpha} (f_{\bm{k^{\prime}},\delta})_i\biggr]\nonumber\\
& \ \times \biggl[V^2 \delta_{\alpha \delta} \delta_{\gamma \beta}+V\alpha_{cf}\bigl\{(\sin{k_y}\delta_{\alpha \delta}\sigma^x_{\gamma \beta}+\sin{k^{\prime}_y}\sigma^x_{\alpha \delta}\delta_{\gamma \beta})\nonumber\\
& \ \ \ +(\sin{k_x}\delta_{\alpha \delta}\sigma^y_{\gamma \beta}+\sin{k^{\prime}_x}\sigma^y_{\alpha \delta}\delta_{\gamma \beta})\bigr\} \nonumber\\
& \ \ \ +\alpha^2_{cf}(\sin{k_x}\sin{k^{\prime}_x}\sigma^y_{\alpha \delta}\sigma^y_{\gamma\beta}+\sin{k_y}\sin{k^{\prime}_y}\sigma^x_{\alpha \delta}\sigma^x_{\gamma\beta}\nonumber\\
& \ \ \ +\sin{k_x}\sin{k^{\prime}_y}\sigma^x_{\alpha \delta}\sigma^y_{\gamma\beta}+\sin{k_y}\sin{k^{\prime}_x}\sigma^y_{\alpha \delta}\sigma^x_{\gamma\beta})\biggr] \label{A1}
\end{alignat}
where we use the following notations: $E_{11}=\bra{1}{\cal H}\ket{1}$, ${\cal H}_{ii}=\bra{i}{\cal H}\ket{i}$, and ${\cal H}_{\bar{i}\bar{i}}=\bra{\bar{i}}{\cal H}\ket{\bar{i}}$.
$(f_{\bm{k}})_i$ is the local component of the $f$-electron at site $i$, i.e. $(f_{\bm{k}})_i=\frac{1}{\sqrt{N}}f_i \exp^{i\bm{k}\cdot \bm{r}_i}$.
If we furthermore assume a dominant scattering at the Fermi energy with $\bm{q}=\bm{k-k^{\prime}}=0$, then
\begin{align}
\thickmuskip=0mu
\medmuskip=0mu
\thinmuskip=0mu
&c^{\dagger}_{\bm{k}\alpha} (f_{\bm{k}\delta})_i \frac{1}{E_{11}\mathalpha{-}{\cal H}_{ii}} (f^{\dagger}_{\bm{k}\gamma})_i c_{\bm{k}\beta}\mathalpha{+} (f^{\dagger}_{\bm{k}\gamma})_i c_{\bm{k}\beta} \frac{1}{E_{11}\mathalpha{-}{\cal H}_{\bar{i}\bar{i}}} c^{\dagger}_{\bm{k}\alpha} (f_{\bm{k}\delta})_i \label{AX} \nonumber\\
\\ 
& \ = \sum_{h=\pm} \biggl[c^{\dagger}_{\bm{k}\alpha} A^{\dagger}_{\delta h} (f_{\bm{k}h})_i \frac{1}{-\frac{U}{2}-h\alpha(\bm{k})} (f^{\dagger}_{\bm{k}h})_i A_{h \gamma} c_{\bm{k}\beta}\nonumber\\
&\ \ \ + (f^{\dagger}_{\bm{k}h})_i A_{h \gamma} c_{\bm{k}\beta} \frac{1}{-\frac{U}{2}+h\alpha(\bm{k})} c^{\dagger}_{\bm{k}\alpha} A^{\dagger}_{\delta h} (f_{\bm{k}h})_i \biggr] \label{A4}\\
& \ \simeq \sum_{h=\pm} \frac{2}{U}(c^{\dagger}_{\bm{k}\alpha} c_{\bm{k}\beta} (f^{\dagger}_{\bm{k}h})_i(f_{\bm{k}h})_i) A^{\dagger}_{\delta h} A_{h \gamma}\label{A5}\\
& \ \ \times\left[\left\{1+\frac{2\alpha(\bm{k})}{U}+(\frac{2\alpha(\bm{k})}{U})^2\right\}+\left\{1-\frac{2\alpha(\bm{k})}{U}+(\frac{2\alpha(\bm{k})}{U})^2\right\}\right]\nonumber\\
& \ =\frac{4}{U}(c^{\dagger}_{\bm{k}\alpha} c_{\bm{k}\beta} (f^{\dagger}_{\bm{k}\gamma})_i(f_{\bm{k}\delta})_i) \left\{1+2(\frac{2\alpha(\bm{k})}{U})^2\right\} \nonumber\\& \ =\frac{4}{NU}(c^{\dagger}_{\bm{k}\alpha} c_{\bm{k}\beta} f^{\dagger}_{i\gamma}f_{i\delta}) \left\{1+2(\frac{2\alpha(\bm{k})}{U})^2\right\}\label{A2}\\
&A=\frac{1}{\sqrt{2}}\left(
    \begin{array}{cc}
      1 & e^{i \theta_{\bm{k}}} \\
      -e^{-i\theta_{\bm{k}}} & 1 \\
      \end{array}
  \right)\\
  &\tan{\theta_{\bm{k}}}=\frac{\sin{k_x}}{\sin{k_y}}\\
&\alpha(\bm{k})= \alpha_{ff} \sqrt{\sin^2{k_x}+\sin^2{k_y}}
\end{align}
$h=\pm 1$ denotes the helical index of the spin polarization. Because we only look at scattering around the unperturbed Fermi surface, we can use $\epsilon_c(\bm{k})=\epsilon_f(\bm{k})=0$. 
We derive Eq(\ref{A5}) from Eq(\ref{A4}) by second order perturbation in $\alpha_{ff}$. 
Furthermore, we can rewrite the hybridization  in Eq. (\ref{A1}) as
\begin{eqnarray}
&=&V^2 \delta_{\alpha \delta} \delta_{\gamma \beta}+V\alpha_{cf}\bigl\{\sin{k_y}(\delta_{\alpha \delta}\sigma^x_{\gamma \beta}+\sigma^x_{\alpha \delta}\delta_{\gamma \beta})\nonumber\\
&&+\sin{k_x}(\delta_{\alpha \delta}\sigma^y_{\gamma \beta}+\sigma^y_{\alpha \delta}\delta_{\gamma \beta})\bigr\}\nonumber\\
&&+\alpha^2_{cf}(\sin^2{k_x}\sigma^y_{\alpha \delta}\sigma^y_{\gamma\beta}+\sin^2{k_y}\sigma^x_{\alpha \delta}\sigma^x_{\gamma\beta}) \nonumber\\
&=&\frac{V^2}{2}(\bm{\sigma}_{\alpha \beta} \cdot \bm{\sigma}_{\gamma \delta}+\delta_{\alpha \beta}\delta_{\gamma \delta})\nonumber\\ 
&&+V\alpha_{cf}\bigl\{\sin{k_y}(\sigma^x_{\alpha\beta}\delta_{\gamma \delta}+\delta_{\alpha\beta}\sigma^x_{\gamma \delta})\nonumber\\
&& \ \ +\sin{k_x}(\sigma^y_{\alpha\beta}\delta_{\gamma \delta}+\delta_{\alpha\beta}\sigma^y_{\gamma \delta})\bigr\}\nonumber\\
&&+ \frac{\alpha^2_{cf}}{2} \bigl\{(\sigma^y_{\alpha\beta}\sigma^y_{\gamma\delta}-\sigma^x_{\alpha\beta}\sigma^x_{\gamma\delta})(\sin^2{k_x}-\sin^2{k_y})\nonumber\\
&&+(\delta_{\alpha\beta}\delta_{\gamma\delta} -\sigma^z_{\alpha\beta}\sigma^z_{\gamma\delta})(\sin^2{k_x}+\sin^2{k_y})\bigr\}\nonumber\\
\label{A3}
\end{eqnarray}
by using
\begin{eqnarray}
\delta_{\alpha \delta} \delta_{\gamma\beta} &=& \frac{1}{2} (\delta_{\alpha\beta} \delta_{\gamma\delta} + \bm{\sigma}_{\alpha\beta}\cdot\bm{\sigma}_{\gamma\delta})\\
\sigma^x_{\alpha \delta} \delta_{\gamma\beta} &=& \sigma^x_{\alpha \mu} \delta_{\mu \delta} \delta_{\gamma\beta}
 =\frac{1}{2} \sigma^x_{\alpha\mu}(\delta_{\mu\beta} \delta_{\gamma\delta} + \bm{\sigma}_{\mu\beta}\cdot\bm{\sigma}_{\gamma\delta})\nonumber\\
 &=&\frac{1}{2} (\sigma^x_{\alpha\beta} \delta_{\gamma\delta} + \delta_{\alpha\beta} \sigma^x_{\gamma\delta}+i(\bm{\sigma}_{\alpha\beta} \times \bm{\sigma}_{\gamma\delta})^x)
\end{eqnarray}
In Eq. (\ref{A3}), terms including $V\alpha_{cf}$ do not correspond to a spin-spin coupling between the $c$- and the $f$-electrons, and can thus be neglected in the Kondo coupling. 
Using Eqs. (\ref{A1}), (\ref{A2}), (\ref{A3}), we can write the effective coupling between the $f$ electron spins and the $c$ electrons as

\begin{eqnarray}
&&{\cal H}^{int}_{eff} = \sum_{i} \sum_{\bm{k} \bm{k^{\prime}}} \frac{1}{N}\exp^{i \bm{q} \cdot \bm{r}_i} J^{\mu\nu}_{\bm{k q}} S^{\mu}_i c^{\dagger}_{\bm{k+q}} \sigma^{\nu} c_{\bm{k}} \\
&&J^{\mu\nu}_{\bm{k 0}}= \left(
    \begin{array}{ccc}
      J_0+J^x_R(\bm{k}) & 0 & 0  \\
      0 & J_0+J^y_R(\bm{k}) & 0 \\
      0 & 0 & J_0-J^z_R(\bm{k}) \\
    \end{array}
  \right)\\
&&\bm{J}_R(\bm{k})=\frac{2\alpha^2_{cf}}{U}(1+8(\frac{\alpha_{ff}(\bm{k})}{U})^2)\left(
    \begin{array}{c}
    -\sin^2{k_y}+\sin^2{k_x}\\
    \sin^2{k_x}-\sin^2{k_y}\\
    \sin^2{k_x}+\sin^2{k_y}\\
    \end{array}
    \right)\nonumber\\
\end{eqnarray}

\section{the poor man's scaling}
For dominant scattering with $\bm{q}=0$, the effective Hamiltonian is described by Eq.(\ref{1}), (\ref{2}), (\ref{3}). The interaction between localized $f$ electrons and $c$ electrons can be written as
\begin{align}
\thickmuskip=0mu
\medmuskip=0mu
\thinmuskip=0mu
&{\cal H}_{int}=\nu \nonumber\\
 &=\frac{1}{N}\sum_{\bm{k},|\bm{q}|<\Delta}c^{\dagger}_{\bm{k+q}\sigma_1}c_{\bm{k}\sigma_2}\biggl[\frac{J^{\prime}_{\bm{q}}}{2}(\sigma^-S^+\mathalpha{+}\sigma^+S^-)\mathalpha{+}J^z_{\bm{q}}\sigma^zS^z\biggr]_{\sigma_1\sigma_2}\nonumber\\
& \ \simeq \frac{1}{N}\sum_{\bm{k},|\bm{q}|<\Delta}c^{\dagger}_{\bm{k+q}\sigma_1}c_{\bm{k}\sigma_2}\biggl[\frac{J^{\prime}_{\bm{0}}}{2}(\sigma^-S^+\mathalpha{+}\sigma^+S^-)\mathalpha{+}J^z_{\bm{0}}\sigma^zS^z\biggr]_{\sigma_1\sigma_2}
\end{align}
where we approximate $J(\bm{q}) \simeq J(0)$.
 Then we can calculate the change of the interaction strengths when the band width of the $c$ electrons shrinks from $E_c$ to $E_c-\Delta E$,
 
 \
 
 \
 
\begin{eqnarray}
&d&\nu 
=\frac{1}{N^2}\sum^{|\epsilon_1|<E_c-\Delta E}_{\bm{k}_1\sigma_1} \ \ \sum^{|\epsilon_2|<E_c-\Delta E}_{\bm{k}_2\sigma_2} \ \ \sum^{E_c-\Delta E<|\epsilon|<E_c}_{\bm{k}\sigma} \nonumber\\
&&\times \biggl[ c^{\dagger}_{\bm{k_2}\sigma_2}c_{\bm{k}\sigma}c^{\dagger}_{\bm{k}\sigma}c_{\bm{k}_1\sigma_1} \frac{1}{\omega-E_c+\epsilon_1} \nonumber\\
&& \ \ \times(\frac{J^{\prime}_{\bm{q_2}}}{2}[S^+\sigma^-_{\sigma_2\sigma}+S^-\sigma^+_{\sigma_2\sigma}]+J^z_{\bm{q_2}}S^z\sigma^z_{\sigma_2\sigma}) \nonumber\\
&& \ \ \times(\frac{J^{\prime}_{\bm{-q_1}}}{2}[S^+\sigma^-_{\sigma\sigma_1}+S^-\sigma^+_{\sigma\sigma_1}]+J^z_{\bm{q_1}}S^z\sigma^z_{\sigma\sigma_1}) \nonumber\\
&&+c^{\dagger}_{\bm{k}\sigma}c_{\bm{k_2}\sigma_2}c^{\dagger}_{\bm{k}_1\sigma_1}c_{\bm{k}\sigma} \frac{1}{\omega-(E_c+\epsilon_1)} \nonumber\\
&& \ \ \times(\frac{J^{\prime}_{\bm{-q_2}}}{2}[S^+\sigma^-_{\sigma\sigma_2}+S^-\sigma^+_{\sigma\sigma_2}]+J^z_{\bm{q_2}}S^z\sigma^z_{\sigma\sigma_2})) \nonumber\\
&& \ \ \times(\frac{J^{\prime}_{\bm{q_1}}}{2}[S^+\sigma^-_{\sigma_1\sigma}+S^-\sigma^+_{\sigma_1\sigma}]+J^z_{\bm{q_1}}S^z\sigma^z_{\sigma_1\sigma})\biggr] ,\label{B2}
\end{eqnarray}
where $\bm{q}_{1(2)}$ means $\bm{k}_{1(2)}-\bm{k}$.
In Eq. (\ref{B2}), the spin exchange interaction term is 
\begin{eqnarray}
&d&\nu_{SEI} \simeq \frac{1}{N}\sum_{\bm{k},|\bm{q}|<\Delta}c^{\dagger}_{\bm{k+q \sigma_1}}c_{\bm{k} \sigma_2}\biggl[-\frac{1}{\omega-E_c+\epsilon_1}\nonumber\\
&&\times\Bigl\{\frac{J^{\prime}_{\bm{0}}J^z_{\bm{0}}}{4}(\sigma^-_{\sigma_1\sigma_2}S^++\sigma^+_{\sigma_1\sigma_2}S^-)+\frac{J^{\prime2}_{\bm{0}}}{2}\sigma^z_{\sigma_1\sigma_2}S^z\Bigr\} \nonumber\\
&&-\frac{1}{\omega-E_c-\epsilon_2}\Bigl\{\frac{J^{\prime}_{\bm{0}}J^z_{\bm{0}}}{4}(\sigma^-_{\sigma_1\sigma_2}S^++\sigma^+_{\sigma_1\sigma_2}S^-)\nonumber\\
&&+\frac{J^{\prime2}_{\bm{0}}}{2}\sigma^z_{\sigma_1\sigma_2}S^z\Bigr\}\biggr]
\end{eqnarray}
Assuming again a dominant scattering around the Fermi surface, we can set $\omega=\epsilon_1=\epsilon_2=0$. We can now derive the renormalization group equations, corresponding to Eqs. (\ref{renorm_z}), (\ref{renorm_p}) in the main text. The details of the derivation of Eq.(\ref{B2}) are written in the textbook by A.C. Hewson. \cite{Hewson} 
\bibliography{paper}

\begin{thebibliography}{37}
\expandafter\ifx\csname natexlab\endcsname\relax\def\natexlab#1{#1}\fi
\expandafter\ifx\csname bibnamefont\endcsname\relax
  \def\bibnamefont#1{#1}\fi
\expandafter\ifx\csname bibfnamefont\endcsname\relax
  \def\bibfnamefont#1{#1}\fi
\expandafter\ifx\csname citenamefont\endcsname\relax
  \def\citenamefont#1{#1}\fi
\expandafter\ifx\csname url\endcsname\relax
  \def\url#1{\texttt{#1}}\fi
\expandafter\ifx\csname urlprefix\endcsname\relax\def\urlprefix{URL }\fi
\providecommand{\bibinfo}[2]{#2}
\providecommand{\eprint}[2][]{\url{#2}}

\bibitem[{\citenamefont{Witczak-Krempa
  et~al.}(2014)\citenamefont{Witczak-Krempa, Chen, Kim, and
  Balents}}]{annurev-conmatphys-020911-125138}
\bibinfo{author}{\bibfnamefont{W.}~\bibnamefont{Witczak-Krempa}},
  \bibinfo{author}{\bibfnamefont{G.}~\bibnamefont{Chen}},
  \bibinfo{author}{\bibfnamefont{Y.~B.} \bibnamefont{Kim}}, \bibnamefont{and}
  \bibinfo{author}{\bibfnamefont{L.}~\bibnamefont{Balents}},
  \bibinfo{journal}{Annual Review of Condensed Matter Physics}
  \textbf{\bibinfo{volume}{5}}, \bibinfo{pages}{57} (\bibinfo{year}{2014}).

\bibitem[{\citenamefont{Peters and Yanase}(2018)}]{PhysRevB.97.115128}
\bibinfo{author}{\bibfnamefont{R.}~\bibnamefont{Peters}} \bibnamefont{and}
  \bibinfo{author}{\bibfnamefont{Y.}~\bibnamefont{Yanase}},
  \bibinfo{journal}{Phys. Rev. B} \textbf{\bibinfo{volume}{97}},
  \bibinfo{pages}{115128} (\bibinfo{year}{2018}),
  \urlprefix\url{https://link.aps.org/doi/10.1103/PhysRevB.97.115128}.

\bibitem[{\citenamefont{Coleman}(2006)}]{coleman2006heavy}
\bibinfo{author}{\bibfnamefont{P.}~\bibnamefont{Coleman}},
  \bibinfo{journal}{arXiv preprint cond-mat/0612006}  (\bibinfo{year}{2006}).

\bibitem[{\citenamefont{Kim et~al.}(2008)\citenamefont{Kim, Jin, Moon, Kim,
  Park, Leem, Yu, Noh, Kim, Oh et~al.}}]{PhysRevLett.101.076402}
\bibinfo{author}{\bibfnamefont{B.~J.} \bibnamefont{Kim}},
  \bibinfo{author}{\bibfnamefont{H.}~\bibnamefont{Jin}},
  \bibinfo{author}{\bibfnamefont{S.~J.} \bibnamefont{Moon}},
  \bibinfo{author}{\bibfnamefont{J.-Y.} \bibnamefont{Kim}},
  \bibinfo{author}{\bibfnamefont{B.-G.} \bibnamefont{Park}},
  \bibinfo{author}{\bibfnamefont{C.~S.} \bibnamefont{Leem}},
  \bibinfo{author}{\bibfnamefont{J.}~\bibnamefont{Yu}},
  \bibinfo{author}{\bibfnamefont{T.~W.} \bibnamefont{Noh}},
  \bibinfo{author}{\bibfnamefont{C.}~\bibnamefont{Kim}},
  \bibinfo{author}{\bibfnamefont{S.-J.} \bibnamefont{Oh}},
  \bibnamefont{et~al.}, \bibinfo{journal}{Phys. Rev. Lett.}
  \textbf{\bibinfo{volume}{101}}, \bibinfo{pages}{076402}
  (\bibinfo{year}{2008}),
  \urlprefix\url{https://link.aps.org/doi/10.1103/PhysRevLett.101.076402}.

\bibitem[{\citenamefont{Neupane et~al.}(2013)\citenamefont{Neupane, Alidoust,
  Xu, Kondo, Ishida, Kim, Liu, Belopolski, Jo, Chang
  et~al.}}]{neupane2013surface}
\bibinfo{author}{\bibfnamefont{M.}~\bibnamefont{Neupane}},
  \bibinfo{author}{\bibfnamefont{N.}~\bibnamefont{Alidoust}},
  \bibinfo{author}{\bibfnamefont{S.}~\bibnamefont{Xu}},
  \bibinfo{author}{\bibfnamefont{T.}~\bibnamefont{Kondo}},
  \bibinfo{author}{\bibfnamefont{Y.}~\bibnamefont{Ishida}},
  \bibinfo{author}{\bibfnamefont{D.-J.} \bibnamefont{Kim}},
  \bibinfo{author}{\bibfnamefont{C.}~\bibnamefont{Liu}},
  \bibinfo{author}{\bibfnamefont{I.}~\bibnamefont{Belopolski}},
  \bibinfo{author}{\bibfnamefont{Y.}~\bibnamefont{Jo}},
  \bibinfo{author}{\bibfnamefont{T.-R.} \bibnamefont{Chang}},
  \bibnamefont{et~al.}, \bibinfo{journal}{Nature communications}
  \textbf{\bibinfo{volume}{4}}, \bibinfo{pages}{2991} (\bibinfo{year}{2013}).

\bibitem[{\citenamefont{Weng et~al.}(2014)\citenamefont{Weng, Zhao, Wang, Fang,
  and Dai}}]{PhysRevLett.112.016403}
\bibinfo{author}{\bibfnamefont{H.}~\bibnamefont{Weng}},
  \bibinfo{author}{\bibfnamefont{J.}~\bibnamefont{Zhao}},
  \bibinfo{author}{\bibfnamefont{Z.}~\bibnamefont{Wang}},
  \bibinfo{author}{\bibfnamefont{Z.}~\bibnamefont{Fang}}, \bibnamefont{and}
  \bibinfo{author}{\bibfnamefont{X.}~\bibnamefont{Dai}},
  \bibinfo{journal}{Phys. Rev. Lett.} \textbf{\bibinfo{volume}{112}},
  \bibinfo{pages}{016403} (\bibinfo{year}{2014}),
  \urlprefix\url{https://link.aps.org/doi/10.1103/PhysRevLett.112.016403}.

\bibitem[{\citenamefont{Lu et~al.}(2013)\citenamefont{Lu, Zhao, Weng, Fang, and
  Dai}}]{PhysRevLett.110.096401}
\bibinfo{author}{\bibfnamefont{F.}~\bibnamefont{Lu}},
  \bibinfo{author}{\bibfnamefont{J.}~\bibnamefont{Zhao}},
  \bibinfo{author}{\bibfnamefont{H.}~\bibnamefont{Weng}},
  \bibinfo{author}{\bibfnamefont{Z.}~\bibnamefont{Fang}}, \bibnamefont{and}
  \bibinfo{author}{\bibfnamefont{X.}~\bibnamefont{Dai}},
  \bibinfo{journal}{Phys. Rev. Lett.} \textbf{\bibinfo{volume}{110}},
  \bibinfo{pages}{096401} (\bibinfo{year}{2013}),
  \urlprefix\url{https://link.aps.org/doi/10.1103/PhysRevLett.110.096401}.

\bibitem[{\citenamefont{Dzero et~al.}(2016)\citenamefont{Dzero, Xia, Galitski,
  and Coleman}}]{doi:10.1146/annurev-conmatphys-031214-014749}
\bibinfo{author}{\bibfnamefont{M.}~\bibnamefont{Dzero}},
  \bibinfo{author}{\bibfnamefont{J.}~\bibnamefont{Xia}},
  \bibinfo{author}{\bibfnamefont{V.}~\bibnamefont{Galitski}}, \bibnamefont{and}
  \bibinfo{author}{\bibfnamefont{P.}~\bibnamefont{Coleman}},
  \bibinfo{journal}{Annual Review of Condensed Matter Physics}
  \textbf{\bibinfo{volume}{7}}, \bibinfo{pages}{249} (\bibinfo{year}{2016}),
  \eprint{https://doi.org/10.1146/annurev-conmatphys-031214-014749},
  \urlprefix\url{https://doi.org/10.1146/annurev-conmatphys-031214-014749}.

\bibitem[{\citenamefont{Dzero et~al.}(2010)\citenamefont{Dzero, Sun, Galitski,
  and Coleman}}]{PhysRevLett.104.106408}
\bibinfo{author}{\bibfnamefont{M.}~\bibnamefont{Dzero}},
  \bibinfo{author}{\bibfnamefont{K.}~\bibnamefont{Sun}},
  \bibinfo{author}{\bibfnamefont{V.}~\bibnamefont{Galitski}}, \bibnamefont{and}
  \bibinfo{author}{\bibfnamefont{P.}~\bibnamefont{Coleman}},
  \bibinfo{journal}{Phys. Rev. Lett.} \textbf{\bibinfo{volume}{104}},
  \bibinfo{pages}{106408} (\bibinfo{year}{2010}),
  \urlprefix\url{https://link.aps.org/doi/10.1103/PhysRevLett.104.106408}.

\bibitem[{\citenamefont{Manchon et~al.}(2015)\citenamefont{Manchon, Koo, Nitta,
  Frolov, and Duine}}]{manchon2015new}
\bibinfo{author}{\bibfnamefont{A.}~\bibnamefont{Manchon}},
  \bibinfo{author}{\bibfnamefont{H.~C.} \bibnamefont{Koo}},
  \bibinfo{author}{\bibfnamefont{J.}~\bibnamefont{Nitta}},
  \bibinfo{author}{\bibfnamefont{S.}~\bibnamefont{Frolov}}, \bibnamefont{and}
  \bibinfo{author}{\bibfnamefont{R.}~\bibnamefont{Duine}},
  \bibinfo{journal}{Nature materials} \textbf{\bibinfo{volume}{14}},
  \bibinfo{pages}{871} (\bibinfo{year}{2015}).

\bibitem[{\citenamefont{Haney et~al.}(2013)\citenamefont{Haney, Lee, Lee,
  Manchon, and Stiles}}]{PhysRevB.87.174411}
\bibinfo{author}{\bibfnamefont{P.~M.} \bibnamefont{Haney}},
  \bibinfo{author}{\bibfnamefont{H.-W.} \bibnamefont{Lee}},
  \bibinfo{author}{\bibfnamefont{K.-J.} \bibnamefont{Lee}},
  \bibinfo{author}{\bibfnamefont{A.}~\bibnamefont{Manchon}}, \bibnamefont{and}
  \bibinfo{author}{\bibfnamefont{M.~D.} \bibnamefont{Stiles}},
  \bibinfo{journal}{Phys. Rev. B} \textbf{\bibinfo{volume}{87}},
  \bibinfo{pages}{174411} (\bibinfo{year}{2013}),
  \urlprefix\url{https://link.aps.org/doi/10.1103/PhysRevB.87.174411}.

\bibitem[{\citenamefont{Jackeli and Khaliullin}(2009)}]{PhysRevLett.102.017205}
\bibinfo{author}{\bibfnamefont{G.}~\bibnamefont{Jackeli}} \bibnamefont{and}
  \bibinfo{author}{\bibfnamefont{G.}~\bibnamefont{Khaliullin}},
  \bibinfo{journal}{Phys. Rev. Lett.} \textbf{\bibinfo{volume}{102}},
  \bibinfo{pages}{017205} (\bibinfo{year}{2009}),
  \urlprefix\url{https://link.aps.org/doi/10.1103/PhysRevLett.102.017205}.

\bibitem[{\citenamefont{Ben~Shalom et~al.}(2010)\citenamefont{Ben~Shalom,
  Sachs, Rakhmilevitch, Palevski, and Dagan}}]{PhysRevLett.104.126802}
\bibinfo{author}{\bibfnamefont{M.}~\bibnamefont{Ben~Shalom}},
  \bibinfo{author}{\bibfnamefont{M.}~\bibnamefont{Sachs}},
  \bibinfo{author}{\bibfnamefont{D.}~\bibnamefont{Rakhmilevitch}},
  \bibinfo{author}{\bibfnamefont{A.}~\bibnamefont{Palevski}}, \bibnamefont{and}
  \bibinfo{author}{\bibfnamefont{Y.}~\bibnamefont{Dagan}},
  \bibinfo{journal}{Phys. Rev. Lett.} \textbf{\bibinfo{volume}{104}},
  \bibinfo{pages}{126802} (\bibinfo{year}{2010}),
  \urlprefix\url{https://link.aps.org/doi/10.1103/PhysRevLett.104.126802}.

\bibitem[{\citenamefont{Fujimoto and Yip}(2012)}]{fujimoto2012aspects}
\bibinfo{author}{\bibfnamefont{S.}~\bibnamefont{Fujimoto}} \bibnamefont{and}
  \bibinfo{author}{\bibfnamefont{S.}~\bibnamefont{Yip}}, in
  \emph{\bibinfo{booktitle}{Non-Centrosymmetric Superconductors}}
  (\bibinfo{publisher}{Springer}, \bibinfo{year}{2012}), pp.
  \bibinfo{pages}{247--266}.

\bibitem[{\citenamefont{Bauer et~al.}(2004)\citenamefont{Bauer, Hilscher,
  Michor, Paul, Scheidt, Gribanov, Seropegin, No\"el, Sigrist, and
  Rogl}}]{PhysRevLett.92.027003}
\bibinfo{author}{\bibfnamefont{E.}~\bibnamefont{Bauer}},
  \bibinfo{author}{\bibfnamefont{G.}~\bibnamefont{Hilscher}},
  \bibinfo{author}{\bibfnamefont{H.}~\bibnamefont{Michor}},
  \bibinfo{author}{\bibfnamefont{C.}~\bibnamefont{Paul}},
  \bibinfo{author}{\bibfnamefont{E.~W.} \bibnamefont{Scheidt}},
  \bibinfo{author}{\bibfnamefont{A.}~\bibnamefont{Gribanov}},
  \bibinfo{author}{\bibfnamefont{Y.}~\bibnamefont{Seropegin}},
  \bibinfo{author}{\bibfnamefont{H.}~\bibnamefont{No\"el}},
  \bibinfo{author}{\bibfnamefont{M.}~\bibnamefont{Sigrist}}, \bibnamefont{and}
  \bibinfo{author}{\bibfnamefont{P.}~\bibnamefont{Rogl}},
  \bibinfo{journal}{Phys. Rev. Lett.} \textbf{\bibinfo{volume}{92}},
  \bibinfo{pages}{027003} (\bibinfo{year}{2004}),
  \urlprefix\url{https://link.aps.org/doi/10.1103/PhysRevLett.92.027003}.

\bibitem[{\citenamefont{Kimura et~al.}(2005)\citenamefont{Kimura, Ito, Saitoh,
  Umeda, Aoki, and Terashima}}]{PhysRevLett.95.247004}
\bibinfo{author}{\bibfnamefont{N.}~\bibnamefont{Kimura}},
  \bibinfo{author}{\bibfnamefont{K.}~\bibnamefont{Ito}},
  \bibinfo{author}{\bibfnamefont{K.}~\bibnamefont{Saitoh}},
  \bibinfo{author}{\bibfnamefont{Y.}~\bibnamefont{Umeda}},
  \bibinfo{author}{\bibfnamefont{H.}~\bibnamefont{Aoki}}, \bibnamefont{and}
  \bibinfo{author}{\bibfnamefont{T.}~\bibnamefont{Terashima}},
  \bibinfo{journal}{Phys. Rev. Lett.} \textbf{\bibinfo{volume}{95}},
  \bibinfo{pages}{247004} (\bibinfo{year}{2005}),
  \urlprefix\url{https://link.aps.org/doi/10.1103/PhysRevLett.95.247004}.

\bibitem[{\citenamefont{Sugitani et~al.}(2006)\citenamefont{Sugitani, Okuda,
  Shishido, Yamada, Thamizhavel, Yamamoto, Matsuda, Haga, Takeuchi, Settai
  et~al.}}]{sugitani2006pressure}
\bibinfo{author}{\bibfnamefont{I.}~\bibnamefont{Sugitani}},
  \bibinfo{author}{\bibfnamefont{Y.}~\bibnamefont{Okuda}},
  \bibinfo{author}{\bibfnamefont{H.}~\bibnamefont{Shishido}},
  \bibinfo{author}{\bibfnamefont{T.}~\bibnamefont{Yamada}},
  \bibinfo{author}{\bibfnamefont{A.}~\bibnamefont{Thamizhavel}},
  \bibinfo{author}{\bibfnamefont{E.}~\bibnamefont{Yamamoto}},
  \bibinfo{author}{\bibfnamefont{T.~D.} \bibnamefont{Matsuda}},
  \bibinfo{author}{\bibfnamefont{Y.}~\bibnamefont{Haga}},
  \bibinfo{author}{\bibfnamefont{T.}~\bibnamefont{Takeuchi}},
  \bibinfo{author}{\bibfnamefont{R.}~\bibnamefont{Settai}},
  \bibnamefont{et~al.}, \bibinfo{journal}{Journal of the Physical Society of
  Japan} \textbf{\bibinfo{volume}{75}}, \bibinfo{pages}{043703}
  (\bibinfo{year}{2006}).

\bibitem[{\citenamefont{Shimozawa et~al.}(2014)\citenamefont{Shimozawa, Goh,
  Endo, Kobayashi, Watashige, Mizukami, Ikeda, Shishido, Yanase, Terashima
  et~al.}}]{PhysRevLett.112.156404}
\bibinfo{author}{\bibfnamefont{M.}~\bibnamefont{Shimozawa}},
  \bibinfo{author}{\bibfnamefont{S.~K.} \bibnamefont{Goh}},
  \bibinfo{author}{\bibfnamefont{R.}~\bibnamefont{Endo}},
  \bibinfo{author}{\bibfnamefont{R.}~\bibnamefont{Kobayashi}},
  \bibinfo{author}{\bibfnamefont{T.}~\bibnamefont{Watashige}},
  \bibinfo{author}{\bibfnamefont{Y.}~\bibnamefont{Mizukami}},
  \bibinfo{author}{\bibfnamefont{H.}~\bibnamefont{Ikeda}},
  \bibinfo{author}{\bibfnamefont{H.}~\bibnamefont{Shishido}},
  \bibinfo{author}{\bibfnamefont{Y.}~\bibnamefont{Yanase}},
  \bibinfo{author}{\bibfnamefont{T.}~\bibnamefont{Terashima}},
  \bibnamefont{et~al.}, \bibinfo{journal}{Phys. Rev. Lett.}
  \textbf{\bibinfo{volume}{112}}, \bibinfo{pages}{156404}
  (\bibinfo{year}{2014}),
  \urlprefix\url{https://link.aps.org/doi/10.1103/PhysRevLett.112.156404}.

\bibitem[{\citenamefont{Petrovic et~al.}(2001)\citenamefont{Petrovic, Pagliuso,
  Hundley, Movshovich, Sarrao, Thompson, Fisk, and
  Monthoux}}]{petrovic2001heavy}
\bibinfo{author}{\bibfnamefont{C.}~\bibnamefont{Petrovic}},
  \bibinfo{author}{\bibfnamefont{P.}~\bibnamefont{Pagliuso}},
  \bibinfo{author}{\bibfnamefont{M.}~\bibnamefont{Hundley}},
  \bibinfo{author}{\bibfnamefont{R.}~\bibnamefont{Movshovich}},
  \bibinfo{author}{\bibfnamefont{J.}~\bibnamefont{Sarrao}},
  \bibinfo{author}{\bibfnamefont{J.}~\bibnamefont{Thompson}},
  \bibinfo{author}{\bibfnamefont{Z.}~\bibnamefont{Fisk}}, \bibnamefont{and}
  \bibinfo{author}{\bibfnamefont{P.}~\bibnamefont{Monthoux}},
  \bibinfo{journal}{Journal of Physics: Condensed Matter}
  \textbf{\bibinfo{volume}{13}}, \bibinfo{pages}{L337} (\bibinfo{year}{2001}).

\bibitem[{\citenamefont{Pe\~na et~al.}(2005)\citenamefont{Pe\~na, Sefrioui,
  Arias, Leon, Santamaria, Martinez, te~Velthuis, and
  Hoffmann}}]{PhysRevLett.94.057002}
\bibinfo{author}{\bibfnamefont{V.}~\bibnamefont{Pe\~na}},
  \bibinfo{author}{\bibfnamefont{Z.}~\bibnamefont{Sefrioui}},
  \bibinfo{author}{\bibfnamefont{D.}~\bibnamefont{Arias}},
  \bibinfo{author}{\bibfnamefont{C.}~\bibnamefont{Leon}},
  \bibinfo{author}{\bibfnamefont{J.}~\bibnamefont{Santamaria}},
  \bibinfo{author}{\bibfnamefont{J.~L.} \bibnamefont{Martinez}},
  \bibinfo{author}{\bibfnamefont{S.~G.~E.} \bibnamefont{te~Velthuis}},
  \bibnamefont{and} \bibinfo{author}{\bibfnamefont{A.}~\bibnamefont{Hoffmann}},
  \bibinfo{journal}{Phys. Rev. Lett.} \textbf{\bibinfo{volume}{94}},
  \bibinfo{pages}{057002} (\bibinfo{year}{2005}),
  \urlprefix\url{https://link.aps.org/doi/10.1103/PhysRevLett.94.057002}.

\bibitem[{\citenamefont{Goh et~al.}(2012)\citenamefont{Goh, Mizukami, Shishido,
  Watanabe, Yasumoto, Shimozawa, Yamashita, Terashima, Yanase, Shibauchi
  et~al.}}]{goh2012anomalous}
\bibinfo{author}{\bibfnamefont{S.}~\bibnamefont{Goh}},
  \bibinfo{author}{\bibfnamefont{Y.}~\bibnamefont{Mizukami}},
  \bibinfo{author}{\bibfnamefont{H.}~\bibnamefont{Shishido}},
  \bibinfo{author}{\bibfnamefont{D.}~\bibnamefont{Watanabe}},
  \bibinfo{author}{\bibfnamefont{S.}~\bibnamefont{Yasumoto}},
  \bibinfo{author}{\bibfnamefont{M.}~\bibnamefont{Shimozawa}},
  \bibinfo{author}{\bibfnamefont{M.}~\bibnamefont{Yamashita}},
  \bibinfo{author}{\bibfnamefont{T.}~\bibnamefont{Terashima}},
  \bibinfo{author}{\bibfnamefont{Y.}~\bibnamefont{Yanase}},
  \bibinfo{author}{\bibfnamefont{T.}~\bibnamefont{Shibauchi}},
  \bibnamefont{et~al.}, \bibinfo{journal}{Physical review letters}
  \textbf{\bibinfo{volume}{109}}, \bibinfo{pages}{157006}
  (\bibinfo{year}{2012}).

\bibitem[{\citenamefont{Peters et~al.}(2016)\citenamefont{Peters, Tada, and
  Kawakami}}]{PhysRevB.94.205142}
\bibinfo{author}{\bibfnamefont{R.}~\bibnamefont{Peters}},
  \bibinfo{author}{\bibfnamefont{Y.}~\bibnamefont{Tada}}, \bibnamefont{and}
  \bibinfo{author}{\bibfnamefont{N.}~\bibnamefont{Kawakami}},
  \bibinfo{journal}{Phys. Rev. B} \textbf{\bibinfo{volume}{94}},
  \bibinfo{pages}{205142} (\bibinfo{year}{2016}),
  \urlprefix\url{https://link.aps.org/doi/10.1103/PhysRevB.94.205142}.

\bibitem[{\citenamefont{Kondo}(1964)}]{kondo1964resistance}
\bibinfo{author}{\bibfnamefont{J.}~\bibnamefont{Kondo}},
  \bibinfo{journal}{Progress of theoretical physics}
  \textbf{\bibinfo{volume}{32}}, \bibinfo{pages}{37} (\bibinfo{year}{1964}).

\bibitem[{\citenamefont{Samokhin et~al.}(2004)\citenamefont{Samokhin, Zijlstra,
  and Bose}}]{PhysRevB.69.094514}
\bibinfo{author}{\bibfnamefont{K.~V.} \bibnamefont{Samokhin}},
  \bibinfo{author}{\bibfnamefont{E.~S.} \bibnamefont{Zijlstra}},
  \bibnamefont{and} \bibinfo{author}{\bibfnamefont{S.~K.} \bibnamefont{Bose}},
  \bibinfo{journal}{Phys. Rev. B} \textbf{\bibinfo{volume}{69}},
  \bibinfo{pages}{094514} (\bibinfo{year}{2004}),
  \urlprefix\url{https://link.aps.org/doi/10.1103/PhysRevB.69.094514}.

\bibitem[{\citenamefont{Zarea et~al.}(2012)\citenamefont{Zarea, Ulloa, and
  Sandler}}]{PhysRevLett.108.046601}
\bibinfo{author}{\bibfnamefont{M.}~\bibnamefont{Zarea}},
  \bibinfo{author}{\bibfnamefont{S.~E.} \bibnamefont{Ulloa}}, \bibnamefont{and}
  \bibinfo{author}{\bibfnamefont{N.}~\bibnamefont{Sandler}},
  \bibinfo{journal}{Phys. Rev. Lett.} \textbf{\bibinfo{volume}{108}},
  \bibinfo{pages}{046601} (\bibinfo{year}{2012}),
  \urlprefix\url{https://link.aps.org/doi/10.1103/PhysRevLett.108.046601}.

\bibitem[{\citenamefont{Chen et~al.}(2016)\citenamefont{Chen, Sun, Tang, and
  Lin}}]{0953-8984-28-39-396005}
\bibinfo{author}{\bibfnamefont{L.}~\bibnamefont{Chen}},
  \bibinfo{author}{\bibfnamefont{J.}~\bibnamefont{Sun}},
  \bibinfo{author}{\bibfnamefont{H.-K.} \bibnamefont{Tang}}, \bibnamefont{and}
  \bibinfo{author}{\bibfnamefont{H.-Q.} \bibnamefont{Lin}},
  \bibinfo{journal}{Journal of Physics: Condensed Matter}
  \textbf{\bibinfo{volume}{28}}, \bibinfo{pages}{396005}
  (\bibinfo{year}{2016}),
  \urlprefix\url{http://stacks.iop.org/0953-8984/28/i=39/a=396005}.

\bibitem[{\citenamefont{Vernek et~al.}(2009)\citenamefont{Vernek, Sandler, and
  Ulloa}}]{vernek2009kondo}
\bibinfo{author}{\bibfnamefont{E.}~\bibnamefont{Vernek}},
  \bibinfo{author}{\bibfnamefont{N.}~\bibnamefont{Sandler}}, \bibnamefont{and}
  \bibinfo{author}{\bibfnamefont{S.}~\bibnamefont{Ulloa}},
  \bibinfo{journal}{Physical Review B} \textbf{\bibinfo{volume}{80}},
  \bibinfo{pages}{041302} (\bibinfo{year}{2009}).

\bibitem[{\citenamefont{Misiorny et~al.}(2011)\citenamefont{Misiorny, Weymann,
  and Barna{\'s}}}]{misiorny2011interplay}
\bibinfo{author}{\bibfnamefont{M.}~\bibnamefont{Misiorny}},
  \bibinfo{author}{\bibfnamefont{I.}~\bibnamefont{Weymann}}, \bibnamefont{and}
  \bibinfo{author}{\bibfnamefont{J.}~\bibnamefont{Barna{\'s}}},
  \bibinfo{journal}{Physical review letters} \textbf{\bibinfo{volume}{106}},
  \bibinfo{pages}{126602} (\bibinfo{year}{2011}).

\bibitem[{\citenamefont{Georges et~al.}(1996)\citenamefont{Georges, Kotliar,
  Krauth, and Rozenberg}}]{RevModPhys.68.13}
\bibinfo{author}{\bibfnamefont{A.}~\bibnamefont{Georges}},
  \bibinfo{author}{\bibfnamefont{G.}~\bibnamefont{Kotliar}},
  \bibinfo{author}{\bibfnamefont{W.}~\bibnamefont{Krauth}}, \bibnamefont{and}
  \bibinfo{author}{\bibfnamefont{M.~J.} \bibnamefont{Rozenberg}},
  \bibinfo{journal}{Rev. Mod. Phys.} \textbf{\bibinfo{volume}{68}},
  \bibinfo{pages}{13} (\bibinfo{year}{1996}),
  \urlprefix\url{https://link.aps.org/doi/10.1103/RevModPhys.68.13}.

\bibitem[{\citenamefont{Bulla et~al.}(2008)\citenamefont{Bulla, Costi, and
  Pruschke}}]{RevModPhys.80.395}
\bibinfo{author}{\bibfnamefont{R.}~\bibnamefont{Bulla}},
  \bibinfo{author}{\bibfnamefont{T.~A.} \bibnamefont{Costi}}, \bibnamefont{and}
  \bibinfo{author}{\bibfnamefont{T.}~\bibnamefont{Pruschke}},
  \bibinfo{journal}{Rev. Mod. Phys.} \textbf{\bibinfo{volume}{80}},
  \bibinfo{pages}{395} (\bibinfo{year}{2008}),
  \urlprefix\url{https://link.aps.org/doi/10.1103/RevModPhys.80.395}.

\bibitem[{\citenamefont{Peters et~al.}(2006)\citenamefont{Peters, Pruschke, and
  Anders}}]{PhysRevB.74.245114}
\bibinfo{author}{\bibfnamefont{R.}~\bibnamefont{Peters}},
  \bibinfo{author}{\bibfnamefont{T.}~\bibnamefont{Pruschke}}, \bibnamefont{and}
  \bibinfo{author}{\bibfnamefont{F.~B.} \bibnamefont{Anders}},
  \bibinfo{journal}{Phys. Rev. B} \textbf{\bibinfo{volume}{74}},
  \bibinfo{pages}{245114} (\bibinfo{year}{2006}),
  \urlprefix\url{https://link.aps.org/doi/10.1103/PhysRevB.74.245114}.

\bibitem[{\citenamefont{Yanase and
  Sigrist}(2008)}]{yanase2008superconductivity}
\bibinfo{author}{\bibfnamefont{Y.}~\bibnamefont{Yanase}} \bibnamefont{and}
  \bibinfo{author}{\bibfnamefont{M.}~\bibnamefont{Sigrist}},
  \bibinfo{journal}{Journal of the Physical Society of Japan}
  \textbf{\bibinfo{volume}{77}}, \bibinfo{pages}{124711}
  (\bibinfo{year}{2008}).

\bibitem[{\citenamefont{Ghaemi and Senthil}(2007)}]{PhysRevB.75.144412}
\bibinfo{author}{\bibfnamefont{P.}~\bibnamefont{Ghaemi}} \bibnamefont{and}
  \bibinfo{author}{\bibfnamefont{T.}~\bibnamefont{Senthil}},
  \bibinfo{journal}{Phys. Rev. B} \textbf{\bibinfo{volume}{75}},
  \bibinfo{pages}{144412} (\bibinfo{year}{2007}),
  \urlprefix\url{https://link.aps.org/doi/10.1103/PhysRevB.75.144412}.

\bibitem[{\citenamefont{Weber and Vojta}(2008)}]{PhysRevB.77.125118}
\bibinfo{author}{\bibfnamefont{H.}~\bibnamefont{Weber}} \bibnamefont{and}
  \bibinfo{author}{\bibfnamefont{M.}~\bibnamefont{Vojta}},
  \bibinfo{journal}{Phys. Rev. B} \textbf{\bibinfo{volume}{77}},
  \bibinfo{pages}{125118} (\bibinfo{year}{2008}),
  \urlprefix\url{https://link.aps.org/doi/10.1103/PhysRevB.77.125118}.

\bibitem[{\citenamefont{Haldane}(1978)}]{PhysRevLett.40.416}
\bibinfo{author}{\bibfnamefont{F.~D.~M.} \bibnamefont{Haldane}},
  \bibinfo{journal}{Phys. Rev. Lett.} \textbf{\bibinfo{volume}{40}},
  \bibinfo{pages}{416} (\bibinfo{year}{1978}),
  \urlprefix\url{https://link.aps.org/doi/10.1103/PhysRevLett.40.416}.

\bibitem[{\citenamefont{Sato and Tsunetsugu}(2016)}]{PhysRevB.94.079907}
\bibinfo{author}{\bibfnamefont{T.}~\bibnamefont{Sato}} \bibnamefont{and}
  \bibinfo{author}{\bibfnamefont{H.}~\bibnamefont{Tsunetsugu}},
  \bibinfo{journal}{Phys. Rev. B} \textbf{\bibinfo{volume}{94}},
  \bibinfo{pages}{079907} (\bibinfo{year}{2016}),
  \urlprefix\url{https://link.aps.org/doi/10.1103/PhysRevB.94.079907}.

\bibitem[{\citenamefont{Hewson}(1997)}]{Hewson}
\bibinfo{author}{\bibfnamefont{A.~C.} \bibnamefont{Hewson}},
  \emph{\bibinfo{title}{The Kondo problem to heavy fermions}},
  vol.~\bibinfo{volume}{2} (\bibinfo{publisher}{Cambridge university press},
  \bibinfo{year}{1997}).

\end{thebibliography}
\end{document}